\PassOptionsToPackage{pdfpagelabels=false}{hyperref} 
\documentclass[fleqn,usenatbib]{mnras}

\usepackage{newtxtext,newtxmath}
\usepackage[T1]{fontenc}

\usepackage{graphicx}	
\usepackage{amsmath}	

\usepackage{amsmath}
\usepackage{setspace}

\title{Characterising the Intracluster Light over the Redshift Range $0.2 < z < 0.8$ in the DES-ACT Overlap}

\author[DES Collaboration]{
\parbox{\textwidth}{
\Large
Jesse~B.~Golden-Marx$^{1}$\thanks{E-mail: jessegm@sjtu.edu.cn},
Y.~Zhang$^{2}$,
R.~L.~C.~Ogando$^{3}$,
S.~Allam$^{4}$,
D.~L.~Tucker$^{4}$,
C.~J.~Miller$^{5,6}$,
M.~Hilton$^{7}$,
B.~Mutlu-Pakdil$^{8}$,
T.~M.~C.~Abbott$^{9}$,
M.~Aguena$^{10}$,
O.~Alves$^{6}$,
F.~Andrade-Oliveira$^{6}$,
J.~Annis$^{4}$,
D.~Bacon$^{11}$,
E.~Bertin$^{12,13}$,
S.~Bocquet$^{14}$,
D.~Brooks$^{15}$,
D.~L.~Burke$^{16,17}$,
A.~Carnero~Rosell$^{18,10,19}$,
M.~Carrasco~Kind$^{20,21}$,
F.~J.~Castander$^{22,23}$,
C.~Conselice$^{24,25}$,
M.~Costanzi$^{26,27,28}$,
L.~N.~da Costa$^{10}$,
M.~E.~S.~Pereira$^{29}$,
J.~De~Vicente$^{30}$,
S.~Desai$^{31}$,
P.~Doel$^{15}$,
S.~Everett$^{32}$,
I.~Ferrero$^{33}$,
B.~Flaugher$^{4}$,
J.~Frieman$^{4,8}$,
J.~Garc\'ia-Bellido$^{34}$,
D.~W.~Gerdes$^{5,6}$,
D.~Gruen$^{14}$,
R.~A.~Gruendl$^{20,21}$,
G.~Gutierrez$^{4}$,
S.~R.~Hinton$^{35}$,
D.~L.~Hollowood$^{36}$,
K.~Honscheid$^{37,38}$,
D.~J.~James$^{39}$,
K.~Kuehn$^{40,41}$,
N.~Kuropatkin$^{4}$,
O.~Lahav$^{15}$,
J.~L.~Marshall$^{2}$,
P.~Melchior$^{42}$,
J. Mena-Fern{\'a}ndez$^{30}$,
R.~Miquel$^{43,44}$,
J.~J.~Mohr$^{45,14}$,
A.~Palmese$^{46}$,
F.~Paz-Chinch\'{o}n$^{20,47}$,
A.~Pieres$^{10,3}$,
A.~A.~Plazas~Malag\'on$^{42}$,
J.~Prat$^{48,8}$,
M.~Raveri$^{49}$,
M.~Rodriguez-Monroy$^{30}$,
A.~K.~Romer$^{50}$,
E.~Sanchez$^{30}$,
V.~Scarpine$^{4}$,
I.~Sevilla-Noarbe,$^{30}$,
C.~Sif\'on,$^{51}$,
M.~Smith,$^{52}$,
E.~Suchyta$^{53}$,
M.~E.~C.~Swanson$^{20}$,
G.~Tarle$^{6}$,
M.~Vincenzi$^{11,52}$,
N.~Weaverdyck$^{6,54}$,
and B.~Yanny$^{4}$
\begin{center} (DES Collaboration) \end{center}
}
\vspace{0.4cm}
\\
}

\date{Accepted XXX. Received YYY; in original form ZZZ}

\pubyear{2022}

\begin{document}
\label{firstpage}
\pagerange{\pageref{firstpage}--\pageref{lastpage}}
\maketitle

\begin{abstract}
We characterise the properties and evolution of Bright Central Galaxies (BCGs) and the surrounding intracluster light (ICL) in galaxy clusters identified in overlapping regions of the Dark Energy Survey and Atacama Cosmology Telescope Survey (DES-ACT), covering the redshift range $0.20<z<0.80$. Using this sample, we measure no change in the ICL's stellar content (between 50-300\,kpc) over this redshift range in clusters with log$_{10}(M_{\rm 200m,SZ}$/M$_{\odot})>$14.4.  We also measure the stellar mass - halo mass (SMHM) relation for the BCG+ICL system and find that the slope, $\beta$, which characterises the dependence of $M_{\rm 200m,SZ}$ on the BCG+ICL stellar mass, increases with radius.  The outskirts are more strongly correlated with the halo than the core, which supports that the BCG+ICL system follows a two-phase growth, where recent growth ($z<2$) occurs beyond the BCG's core.  Additionally, we compare our observed SMHM relation results to the IllustrisTNG 300-1 cosmological hydrodynamic simulations and find moderate qualitative agreement in the amount of diffuse light.  However, the SMHM relation's slope is steeper in TNG300-1 and the intrinsic scatter is lower, likely from the absence of projection effects in TNG300-1.  Additionally, we find that the ICL exhibits a colour gradient such that the outskirts are bluer than the core.  Moreover, for the lower halo mass clusters (log$_{10}(M_{\rm 200m,SZ}$/M$_{\odot})<$14.59 ), we detect a modest change in the colour gradient's slope with lookback time, which combined with the absence of stellar mass growth may suggest that lower mass clusters have been involved in growth via tidal stripping more recently than their higher mass counterparts. 
\end{abstract}
\begin{keywords}
galaxies: clusters: general -- galaxies: elliptical and lenticular, cD -- galaxies: evolution 
\end{keywords}

\section{Introduction}
\label{sec:intro}
Bright Central Galaxies (BCGs) sit at the center of galaxy clusters.  This position ties their formation and properties to the cluster's underlying dark matter halo.  One key characteristic of BCGs is that they are observed to be surrounded by a halo or envelope of diffuse light \citep[e.g.,][]{zwi51,zwi52,mat64,mor65}, commonly referred to as intracluster light (ICL).  Despite early observations of the ICL, it wasn't until the advent of CCDs and higher resolution optical observations that it became possible to characterise and measure the properties of the ICL.

Observations using stacked ICL measurements \citep{zib05,zha18,che22} find that this diffuse envelope extends out to hundreds of kpc and possibly even out to Mpc scales.  However, although high resolution and large field of view images allow us to detect the ICL out to large radii, it becomes impossible to do so observationally at small radii, because it remains exceedingly difficult to distinguish between the light and stars associated with the BCG from that of the ICL.  Therefore, in this analysis, we do not attempt to disentangle these regimes and instead focus on the BCG+ICL system.  Moreover, we follow the definition introduced in \citet{pil18} and used in \citet{zha18} and \citet{san21} and define the ICL as all light beyond a fixed physical aperture of 50\,kpc.  

One benefit of focusing on the BCG+ICL system is that it follows the general consensus that the formation and evolution of the BCG and ICL are intrinsically linked \citep[e.g.,][]{mih05,mur07,pur07,puc10,rud11,con14,con18,con19,bur15,dem15,dem18,gro17,mor17,mon18,che22}.  The ICL is composed of unbound stars, many of which formed within galaxies that have interacted with the BCG and become part of the ICL as a result of such interactions through processes including galaxy disruption \citep{guo11}, tidal or stellar stripping of satellite galaxies \citep[e.g.,][]{gal72,dem15,dem18,mor17,mon18}, and relaxation following galaxy mergers \citep{mur07,mor17}.  The latter two methods are particularly key to our current understanding of BCG formation.  Based on observations and simulations, BCGs are thought to form as a result of a two-phase formation scenario \citep{ose10,van2010}, where the core of the BCG forms at high redshifts, $z > 2$, and the outskirts of the BCG, which include the ICL, continue to grow as a result of major or minor mergers or stellar stripping.  This formation process makes the ICL's formation intrinsically linked to the hierarchical growth of the BCG. 

Since it is challenging to observationally measure the ICL as a result of its diffuse nature and low surface brightness, often near the background level, much work has focused on lower redshift ($z < 0.1$) observations \citep[e.g.,][]{mih17}.  Despite these difficulties, some analyses have extended our understanding of the ICL out to intermediate redshifts \citep[$0.2 < z < 0.5$;][]{dem15,mon18,zha18,san21, fur21, che22}, and even to higher redshifts \citep[$z\approx 1$ ][]{bur12,bur15,dem18,ko18}.  Since it is viable to study the ICL across redshift, some studies \citep[e.g.,][]{fur21} have investigated the redshift evolution of ICL properties. For example, \citet{fur21} use a sample of 18 X-ray selected clusters with Hyper Suprime Cam Subaru Strategic Program observations and find that over the redshift range $0.1 < z < 0.5$ the ICL grows by a factor of 2-4 when measured out to $R_{500}$.

The plethora of recent observations highlight some key characteristics of the ICL, particularly the stellar content and spatial distribution.  Since BCG's account for a significant fraction of the cluster's total light \citep[e.g.,][]{sch86,jon00,lin04,ber07,lau07,von07,agu11,bro11,pro11,har12}, it follows that the BCG+ICL system does as well, with estimates between 10\% and 50\% \citep{zib05,gon07,bur15,mon18}.  Moreover, previous works have included not only the BCG's core, but also the surrounding ICL, to further characterise the galaxy-dark matter halo connection. For example, prior analyses \citep{zib05,hua18,dem18,gol19,zha19,klu20,san21,hua21} have detected a Stellar Mass - Halo Mass (SMHM) correlation associated with either the outer envelopes of the BCG or the ICL.  The correlation between the ICL and the underlying host halo is further supported by observational measurements of the velocity dispersion.  Many observations of massive early type galaxies have found that in the outskirts of the BCG+ICL system, the velocity dispersion of the stars rises to match the underlying velocity dispersion of the cluster, suggesting that the ICL may trace the cluster's gravitational potential \citep[e.g.,][]{fab77,dre79,kel02,ben15,vea18,edw20,gu20}. This concept has been further studied in recent years.  Some analyses \citep[e.g.,][]{mon19} have found that the spatial distribution of the ICL matches the shape of the cluster's underlying dark matter distribution, which underscores that although the ICL surrounds the BCG and may be tied to the BCG's formation, the stars in the ICL are not bound to the BCG, but are instead bound to the cluster's dark matter halo.  In addition to stellar mass, the colour of the ICL has also been used to understand the growth history of the ICL. Recent detections of a colour gradient \citep{dem15,dem18,mor17} in ICL observations suggest that the ICL forms primarily through the tidal stripping of satellite galaxies.

Here we present measurements of the stellar content, as well as the correlation with halo mass and BCG colour as a function of both radial extent and redshift performed for the first time using a large sample of high quality observations covering a large volume.  To do so, we utilise the Dark Energy Survey \citep[DES;][]{DES05} - Atacama Cosmology Telescope \citep[ACT;][]{hil20} overlap sample of clusters covering the redshift range $0.2 < z < 0.8$ to characterise the colour and stellar mass content of the ICL.

The remainder of this paper is outlined as follows.  In Section~\ref{sec:data}, we define our observational sample, the DES-ACT sample, as well as our data reduction methodologies.  In Section~\ref{sec:measurements}, we present the measurements of the stellar mass content of the ICL.  In Section~\ref{sec:SMHM} we present our analysis on the SMHM relation for the BCG+ICL system.  In Section~\ref{sec:TNG}, we describe our measurement of the BCG+ICL SMHM relation in the TNG300-1 simulation. In Section~\ref{sec:colour}, we present results relating to the colour and radial aperture of the ICL.  Finally, we conclude in Section~\ref{sec:conclusion}.  

Except for the simulated TNG300-1 data, in which the cosmological parameters are predefined ($\Omega_{M}$=0.3089, $\Omega_{\Lambda}$=0.6911, H$_{0}$=100~$h$~km/s/Mpc with $h$=0.6774), throughout this analysis, we assume a flat $\Lambda$CDM universe, with $\Omega_{M}$=0.30, $\Omega_{\Lambda}$=0.70, H$_{0}$=100~$h$~km/s/Mpc with $h$=0.7.

\section{data}
\label{sec:data}
\subsection{DES-ACT data}
\label{subsec:DES-ACT}
In this analysis, we aim to determine how the ICL (as well as stellar content within galaxy clusters) evolves over cosmic time.  As a result of the difficulties (discussed in Section~\ref{subsec:ICL_measurement}) with measuring the diffuse ICL, it is paramount that such an analysis be done on a single sample measured across a large redshift range, rather than a joint sample, where differences in the surface brightness limits and measurement uncertainties need to be accounted for \citep[e.g.,][]{gol22}.  Therefore, we use the galaxy clusters detected via the Sunyaev-Zel'dovich (SZ) effect \citep{sun70,sun72} from the Atacama Cosmology Telescope Data Release 5 (DR5), presented in \citet{hil20}.  Clusters detected via the thermal SZ effect yield a redshift independent, but mass limited cluster sample, unlike samples of optically detected clusters.  In total, the ACT sample contains 4195 optically-confirmed SZ-detected clusters, identified using the 98 and 150 GHz channels, covering the redshift range of $0.04 < z < 1.91$.  For a complete description of the ACT-SZ cluster catalog, see \citet{hil20}.  Here we summarise the necessary details from the ACT-SZ catalog.  Each cluster in the sample has an SZ-estimated halo mass, $M_{\rm 200m,SZ}$, and an associated uncertainty.  The ACT DR5 cluster catalog contains mass estimates derived from the measured SZ signal for each cluster, assuming the \citet{Arn10} scaling relation. As a sample, the catalog is found to be more than 90$\%$ complete for halo masses greater than $3.8\times 10^{14}$ M$_{\odot}$ ( log$_{10}(M_{500c,SZ}/$M$_{\odot}) > 14.58$; see Figure 7 in \citet{hil20}) which yields a large sample of the highest mass clusters in the universe spread across a large redshift range.

\begin{figure}
    \centering
    \includegraphics[width=\columnwidth]{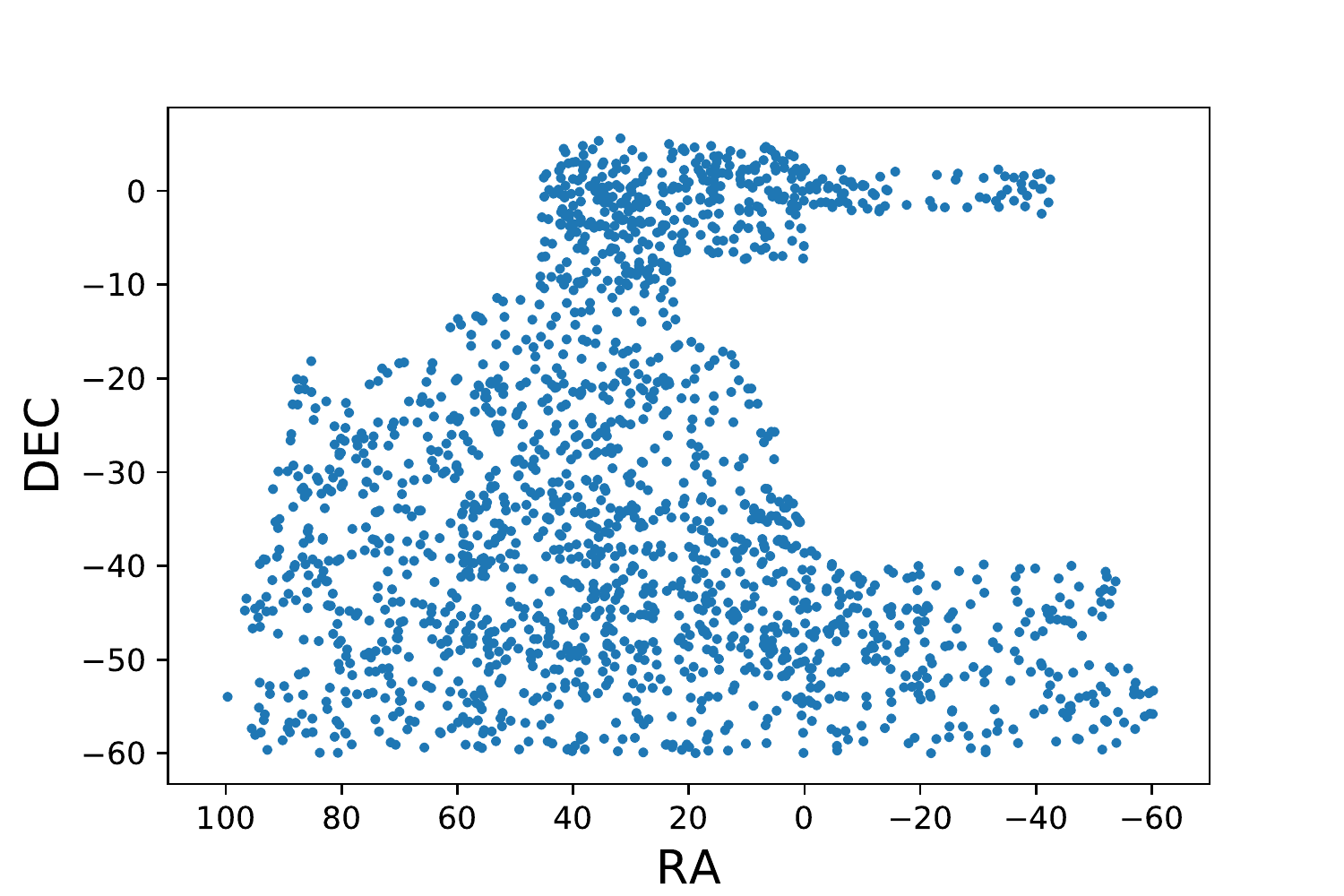}
    \caption{The spatial distribution of the 1455 clusters in the overlap region between DESY3 and ACT DR5.}
    \label{fig:ACT_DES}
\end{figure}

To study the BCG+ICL system and take advantage of the ACT catalogs, we use the 4566 deg$^2$ overlap between the Dark Energy Survey Year 3 (DES Y3) v6.4.22 redMaPPer and the ACT catalogs, which includes 1455 clusters.  The spatial distribution of these clusters is shown in Figure~\ref{fig:ACT_DES}. 
We note that each DES-ACT cluster has been confirmed using the redMaPPer optical cluster finder \citep{ryk16}, which relies on overdensities in the griz colour space to identify clusters \citep{ryk14} out to $z=0.9$ in DESY3 data.  As a result of the cross-identification in both DESY3 and ACT each cluster has a redMapper estimated photometric redshift, which we utilise when we estimate the radial aperture measurements for the ICL.  Moreover, by using the DES-ACT data, we take advantage of the deep DES photometry, as done in \citet{gol22}, to measure the combined BCG+ICL light profiles out to large radii.  We note that here, we only analyse data out to redshift $z=0.8$ instead of $z=0.9$, because beyond $z=0.8$, the cluster identification is incomplete. This redshift cut reduces the number of clusters from 1455 to 1199.          

\subsection{DES Y6 data}
\label{subsec:DESy6}
The Dark Energy Survey is a wide-area multi-band photometric survey that used the Victor M. Blanco 4-m telescope to observe $\approx$ 5,000 square degrees in five bands ($g, r, i, z, Y$) from 2012 (Science Verification) until early 2019 (completing Year 6), co-adding several layers of single-epoch images into exquisite, deep, co-add images. DECam, with a large 2.2 degree field-of-view, a CCD mosaic of 570-megapixels, and a low-noise electronic readout system, is an ideal instrument to study extended diffuse sources in a single field of view, like the ICL in galaxy clusters.  

The data used here, both images and source catalogs, come from the co-addition of all 6 years of DES data and are available as part of Data Release 2 \citep[][]{DESDR2}. This includes deeper and more uniform co-added images with better background subtraction than previous data releases. An important consequence to our work is better masking of faint galaxies close to the diffuse light, which enhances the ability to distinguish between the diffuse and total light within the cluster, which includes cluster member galaxies and the ICL.

To achieve the improvements described in \citet{DESDR2}, additional exposures and tilings covering the entire DES footprint are used to create the final image.  For such images, only single-epoch background subtraction is performed, rather than a ``global" Swarp \citep{2010ascl.soft10068B} background subtraction, which can lead to the appearance of artefacts.  For this analysis, the detection images are an average of the $riz$ images and the detection threshold is lower in the DESY6 data (5$\sigma$) than what was used in previous releases (10$\sigma$).

\subsection{Measuring the ICL}
\label{subsec:ICL_measurement}
The BCG and ICL profiles are measured using similar procedures in \citet{zha18}, \citet{san21}, and \citet{gol22}. For each galaxy cluster, centered on the redMaPPer central galaxy, we adopt the following procedure, 
\begin{itemize}
    \item We query the DES database and identify images that cover a {0.15 deg $\times$ 0.15 deg} region around the BCG. These images are stacked (using their mean pixel values) to create a coadd image covering a {0.15 deg $\times$ 0.15 deg} region centered on the BCG. These images have only gone through a global background subtraction process \citep[estimated from the whole Field-of-View, about 2.5 $\mathrm{deg}^2$ for single exposure images;][]{2017PASP..129k4502B}. No local sky background subtraction process, either through Source Extractor \citep{1996A&AS..117..393B} or Swarp \citep{2010ascl.soft10068B}, is applied. 
    
    \item If the BCG is surrounded by either a bright star or a bright foreground galaxy, we remove that cluster from our sample, since we are unable to adequately measure the ICL in these poor regions.  As a result, this reduces the number of clusters in our analysis sample from 1199 to 1137.  
    
    \item Centered on the BCG, we query the DES database for all objects covering a 0.2 deg $\times$ 0.2 deg area of the region surrounding the central galaxy.  We assume the most probable central galaxy from redMaPPer is the BCG.  While miscentering does exist in the redMaPPer catalog (see \citet{zha19} for a complete discussion), we do not correct for this here and assume that miscentering may yield added noise within our measurements.  Using the Kron radius \citep{Kron80} measurements as well as ellipticity and orientation measurements of the redMaPPer BCGs, we mask an elliptical region covering those objects with a semi-major axis of 3.5 Kron radius. There are exclusions to this masking procedure: the BCG is not masked, to preserve the measurements of the BCG's light profile; galaxies fainter than a magnitude threshold, set according to the galaxy cluster's photometric redshifts, are also not masked to ensure similar masking luminosity ranges for the all of the clusters.  Here, we use an M$_{*}$+2 masking threshold in the DES i-band, given in Table~\ref{tab:limits}.  
    \begin{table}{cc}
    \centering
    \caption{Masking Magnitude Threshold}
    \begin{tabular}{cc}
    \hline
    redshift range & DES i-band M$_{*}$+2 masking threshold \\
    \hline 
    0.05 $<$ z $<$ 0.29 & 20.191 \\
    0.29 $<$ z $<$ 0.40 & 21.163 \\
    0.40 $<$ z $<$ 0.50 & 21.730 \\
    0.50 $<$ z $<$ 0.58 & 22.229 \\
    0.58 $<$ z $<$ 0.66 & 22.565 \\
    0.66 $<$ z $<$ 0.80 & 22.975 \\
    \hline
    \end{tabular}
    \label{tab:limits}
    \end{table}
    
    \item After masking, we measure the surface brightness profiles of each galaxy cluster from those unmasked regions. Centered on the central galaxy, we measure the average galaxy cluster's surface brightness values (mean value) in radial annuli, and those measurements become the surface brightness radial profile for each cluster, containing primarily the light from the cluster's BCG and the intracluster light component as well as some background light. Masked regions are excluded from the measurement process.  These masked measurements are used to measure the BCG+ICL system for each galaxy cluster. 
    
    \item {\it Including the masked regions}, we also made another version of the surface brightness measurements described in the above step. These measurements contain light coming from all of the cluster galaxies, background galaxies, as well as the light from the cluster central galaxies and the intracluster light.  These unmasked measurements are used to measure the total stellar content of the galaxy cluster and referred to as the total light in this analysis.
    
    \item The above surface brightness measurements also contain the background residual light as well as light contributions from foreground and background objects. For each of the surface brightness radial measurements (each cluster, and each of its unmasked and masked measurements), we take their values at the radial range beyond  500\,kpc as the ``background" value, and subtract this ``background" value from the corresponding measurements.  The choice of 500\,kpc is discussed in the Appendix.
\end{itemize}

\section{Measurements}
\label{sec:measurements}
\subsection{Stellar Mass - Halo Mass Parameter Measurements}
\label{subsec:SMHMparams}
Along with measuring the ICL associated with each cluster, one central tenant of this analysis is studying the SMHM relation and determining the impact of incorporating the ICL.  As such, we require an understanding of our stellar and halo mass measurements.  As discussed in Section~\ref{subsec:DES-ACT}, for each cluster, the halo mass is $M_{\rm 200m,SZ}$ with their individually estimated uncertainties.  

As described in Section~\ref{subsec:ICL_measurement}, for each cluster, we determine the apparent magnitude of the BCG+ICL system within each discrete radial regime by measuring the BCG+ICL system's light profile and then integrating that profile over the different physical apertures.  Using these magnitudes, we estimate the stellar mass using the EzGal \citep{man12} SED modelling software.  In this approach, we assume a passively evolving spectral model because our available photometry limits us from statistically constraining additional parameters, such as burst times and formation epochs.  For our mass estimate, we use the same parameters as \citet{gol19} and \citet{gol22}; we assume a \citet{bru03} stellar population synthesis model, a \citet{sal55} Initial Mass Funtion (IMF), a formation redshift of $z=4.9$, and a metallicity of 0.008 (66\%~Z$_{\odot}$).  We note that our stellar mass estimate is independent of formation redshift.  As in \citet{gol22}, we assume a subsolar metallicity for each BCG+ICL system since observations find that light from the outskirts of the BCG envelope, including the ICL are characterised by a subsolar metallicity \citep{mon18} and that the metallicity of early type galaxies decreases radially outward \citep{mcd2015}.  Using these assumptions, we estimate the stellar mass of the BCG+ICL system using the DES i-band magnitude. 

We note that this approach differs from what was used in \citet{gol22}, where a colour estimated stellar mass was calculated.  However, as discussed in Section~\ref{sec:colour}, the BCG+ICL system's colour varies radially, which would add noise to our mass estimate. Since the magnitudes were estimated in the same manner as in \citet{gol22} and the stellar masses were also estimated using EzGal, we assume the same uncertainty, 0.06 dex, in the BCG+ICL stellar mass as used in \citet{gol22}.  This may be a lower limit on the estimated uncertainty in stellar mass.  However, if this is the case, this would only change the estimated value of the intrinsic scatter ($\sigma_{int}$) in stellar mass at fixed halo mass, the remaining parameters would remain statistically consistent.  Additionally, we note that were we to select a \citet{cha03} IMF, this would uniformly shift our stellar mass estimates down by $\approx$~0.25 dex.  However, this would only result in a change to the amplitude of the SMHM relation, so we don't include it in our uncertainty.    

\subsection{Volume Complete Sample}
\label{subsec:sample}
One of the primary goals of this analysis is to identify evolution in the stellar content of the ICL over the redshift range $0.2 < z < 0.8$ as well as the SMHM relation.  One approach to answering this question is to look at whether the stellar mass of the ICL changes with redshift.  We note that we apply a lower redshift limit of $z=0.2$ to maintain uniform photometric coverage in the DES i- and r-bands across our redshfit range.  Applying this lower limit reduces the number of clusters from 1137 to 1069.  

\begin{figure}
    \centering
    \includegraphics[width=\columnwidth]{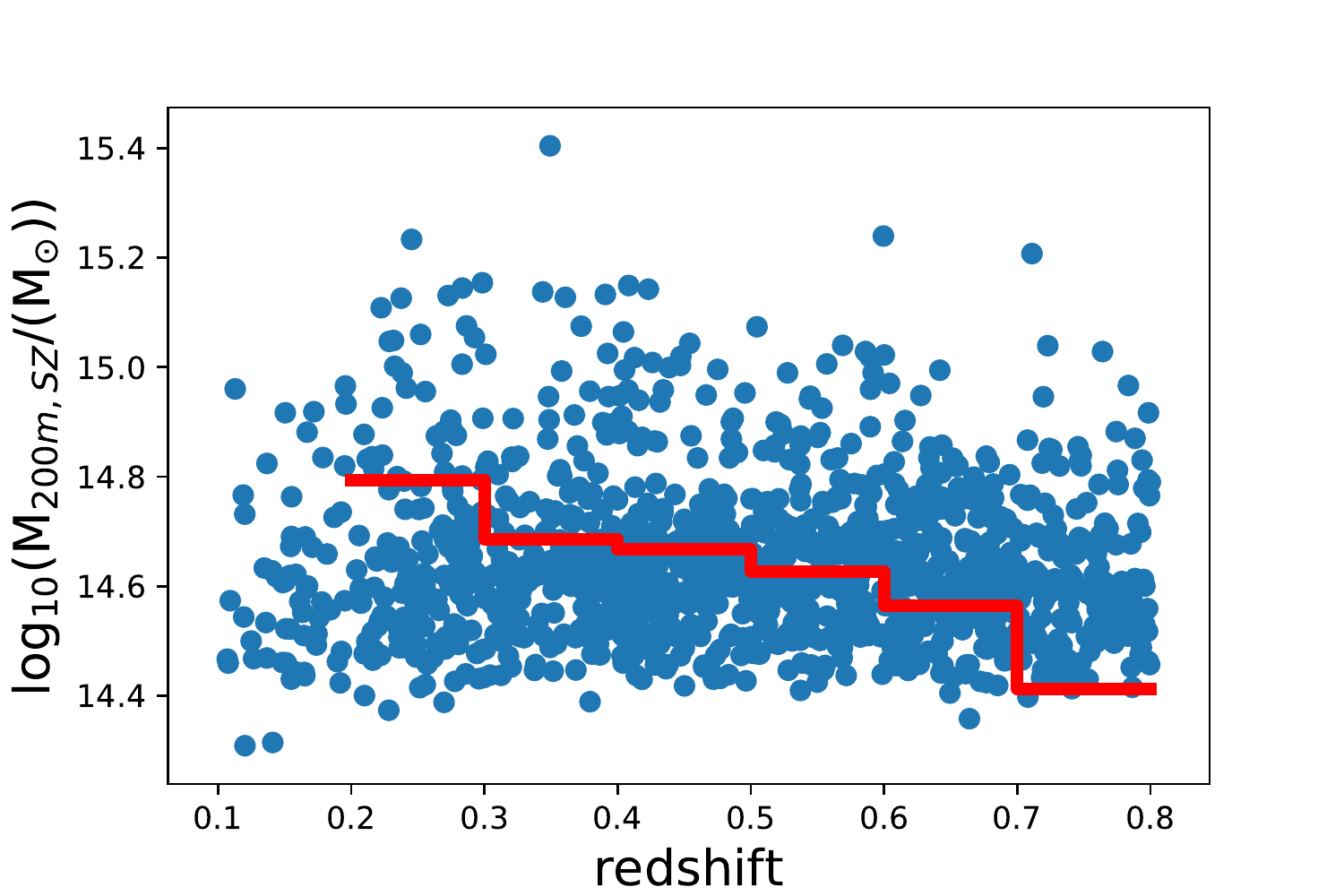}
    \caption{The distribution of $M_{\rm 200m,SZ}$ for the 1199 DESY3-ACT clusters as a function of redshift.  The red line represents the completeness cuts used to eliminate halo mass evolution and differences in the volume observed.  The blue data above the red line is used when studying the evolution of the total stellar mass within the ICL and the SMHM relation.}
    \label{fig:Mhalo_z_cut}
\end{figure}

We aim to first ensure that in each redshift range, we are comparing similar populations of galaxy clusters. Since cluster halo masses may evolve with redshift and the ACT cluster samples have a similar mass threshold across the entire redshift range, we apply a selection cut so that the volume density of the galaxy clusters remains constant across the redshift range. For this procedure, we bin the data in 6 redshift bins from 0.2 to 0.8 and treat the bin with $0.7 < z < 0.8$ as the pivotal bin.  We then determine how much bigger in cosmic volume each of the lower redshift bins are compared to the pivotal bin.  Using this volume estimate, we determine a new mass threshold for each lower redshift bin so that the volume density of the selected clusters stays the same across all redshift bins.  As a result of applying this fixed-density selection, we remove many of the lower M$_{\rm 200m,SZ}$ clusters at lower redshifts and reduce the available number of clusters in our analysis from 1069 to 511.  We visually convey this in Figure~\ref{fig:Mhalo_z_cut}, where all clusters below the red line are removed from our sample.  As previously noted, prior to this selection cut, the mass threshold of clusters shown in Figure~\ref{fig:Mhalo_z_cut} is almost uniform with redshift, which results from the cluster's SZ identification.

\subsection{Stellar Content of the ICL}
\label{subsec:ICL measurement}
For the purposes of this analysis, we treat the radial regime between 50\,kpc and 300\,kpc from the center of the BCG as the ICL.  This radial range is similar to the BCG+ICL ``transitional region'', introduced by \citet{che22}, which consists of the region beyond the BCG's core where the gravitational influence of the BCG continues to strongly impact growth, making it an ideal regime to study the BCG+ICL stellar content.  

\begin{figure*}
    \centering
    \includegraphics[width=18cm]{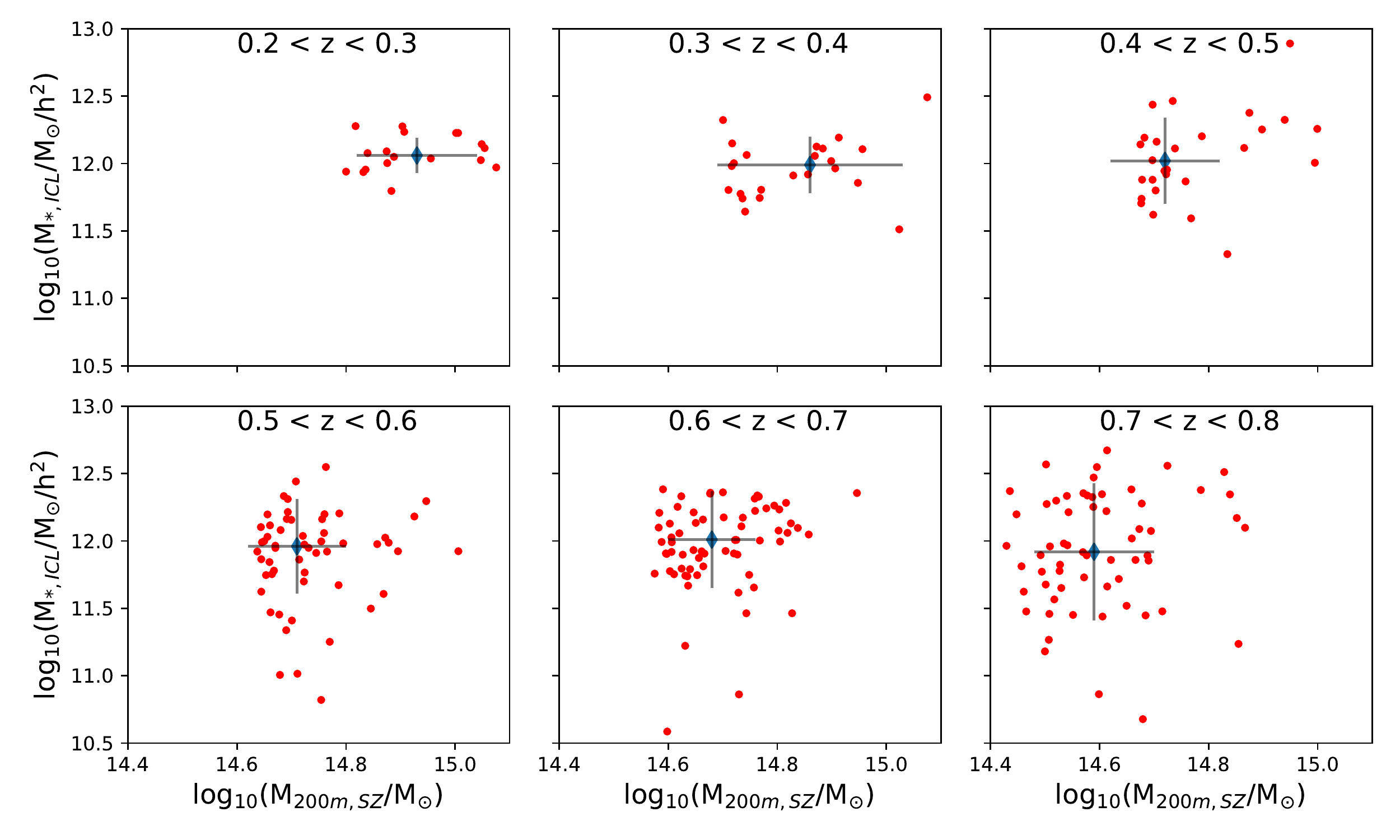}
    \caption{The total stellar mass contained within the ICL between 50\,kpc and 300\,kpc as a function of $M_{\rm 200m,SZ}$ for a volume complete halo mass limited sample of the DES-ACT clusters.  The blue diamonds and gray error bars represent the median and standard deviation of the stellar mass and M$_{\rm 200m,SZ}$.  We see that the median total stellar mass estimate remains relatively constant over this range in redshift.}
    \label{fig:ICL_Mstar}
\end{figure*}

Perhaps the simplest way to estimate the evolution of the stellar mass of the ICL is to illustrate how it varies with $M_{\rm 200m,SZ}$ across redshift space for our volume complete sample.  Figure~\ref{fig:ICL_Mstar} shows the ICL's stellar mass, based on the M$_{*}$+2 masking limit, given in Table~\ref{tab:limits}, as a function of $M_{\rm 200m,SZ}$.  As a result of strict masking, we are only able to measure the ICL in 260 clusters.
We note that as shown by the median values (the blue diamonds) in Figure~\ref{fig:ICL_Mstar} we are unable to detect any evolution in the ICL's median stellar mass with redshift. Although $<M_{\rm 200m,SZ}>$ decreases with redshift for this volume complete sample, the median stellar mass remains constant over the redshift range $0.2 < z < 0.8$.  

Based on previous results \citep[e.g.,][]{fur21}, the absence of redshift evolution is somewhat surprising.  However, this result may be due to the our choice to truncate the ICL at 300kpc (as opposed to R$_{500}$, which  \citet{fur21} use), because the recent growth of the ICL may occur at larger radii. We note that using the DESY6 data, the ICL becomes too noisy beyond 300kpc to measure for individual clusters and can only be measured using a stacking analysis at such large radii.  Additionally, the limited range in $M_{\rm 200m,SZ}$ of the volume complete sample (in each redshift slice) may also contribute to the lack of evolution.  The lack of change in the stellar content of the ICL for this volume complete sample illustrates that we are unable to detect any noticeable changes in the stellar mass of the ICL within the radial range of 50\,kpc to 300\,kpc over the redshift range $0.2 < z < 0.8$.  As a result of the absence of redshift evolution, we do not model the parameters of the SMHM relation to evolve with redshift. 

\section{The Observed SMHM Relation}
\label{sec:SMHM}
The SMHM relation for clusters, compares the stellar mass within the BCG to the host halo mass.  This is one of the most commonly used scaling relations to characterise the galaxy-dark matter halo connection \citep[e.g.,][]{lin04,beh13,mos13,zu16,kra14,gol18,gol19,gol22}.  For clusters, this power law relation is particularly useful because it is characterised by a small intrinsic scatter \citep[$\sigma_{int} \approx 0.15 $dex; e.g.,][]{zu16,pil17, kra14,gol18} in stellar mass at fixed halo mass for both observations and simulations.  

As previously noted, recent observations \citep[e.g.,][]{mon19} suggest that the ICL distribution traces the dark matter distribution of the underlying host halo and earlier results have identified qualitative correlations between the ICL's stellar mass and the halo mass \citep[e.g.,][]{san21,hua21}.  Since the correlation between ICL stellar mass and halo mass has not been quantified, it remains unclear how the SMHM relation changes when the light from the ICL is incorporated.  As discussed in Section~\ref{sec:intro}, the slope of the SMHM relation increases as the outer radii within which stellar mass is measured increases as a result of the BCG+ICL system's inside-out growth \citep{ose10,van2010}.  Moreover, the outer envelope should be more tightly correlated with halo mass than the core because the stellar content within the outer envelope reflects the recent merger growth of the BCG+ICL system and host dark matter halo.  \citet{gol19} find that the slope of the SMHM relation for the BCG+ICL system should increase until at least a radius of $\approx$50\,kpc.  Beyond this radius, the authors find that the unstacked SDSS data becomes background limited.  However, since we use the deeper DES data, it's possible that the increasing trend between radius and slope may extend to radii beyond 50\,kpc. 

We also note that in \citet{gol19}, when the SMHM relation (Equation~\ref{eq:SMHM}) was measured as a function of radius, the change in slope was modest (0.1 between 20\,kpc and 100\,kpc).  In contrast, \citet{gol19} measure an increase in the SMHM relation's slope of 0.25 over the same radial range when incorporating the magnitude gap, the difference in brightness between the BCG and 4th brightest member galaxy within 0.5R$_{200}$, as a third parameter in the SMHM relation.  In both scenarios, the slope of the SMHM relation asymptotes at a radius around 50-60\,kpc.  Here we choose not to investigate the magnitude gap because doing so may further reduce the available sample, since to measure the magnitude gap we need three high probability members.  Additionally, the correlation between the magnitude gap and ICL remains unclear and will be analysed in a future work.  We note that since we are not incorporating the magnitude gap, as done in \citet{gol19}, some of the impact of the change in radial aperture may be slightly diminished. 

\subsection{The BCG+ICL and Total Cluster Light SMHM relations}
The SMHM relation is quantified using Equation~(\ref{eq:SMHM}),
\begin{equation}
\label{eq:SMHM}
    log_{10}(M_{*}/\rm{M}_{\odot})=\textit{N}(\alpha + \beta*log_{10}(M_{\rm 200m,SZ}/\rm{M}_{\odot}),\sigma_{int}^2)
\end{equation} 
where $\alpha$ is the offset, $\beta$ is the slope of the SMHM relation, and $\sigma_{int}$ is the intrinsic scatter in stellar mass at fixed $M_{\rm 200m,SZ}$.  We then measure how the correlation between $\beta$, $\sigma_{int}$, and radius evolves out to the ICL, using fixed physical radii to define the outer boundary.  For this analysis, we use the regions of 0-10\,kpc, 0-30\,kpc, 0-50\,kpc, 0-100\,kpc, 0-200\,kpc, and 0-300\,kpc. 

\begin{figure*}
    \centering
    \includegraphics[width=18cm]{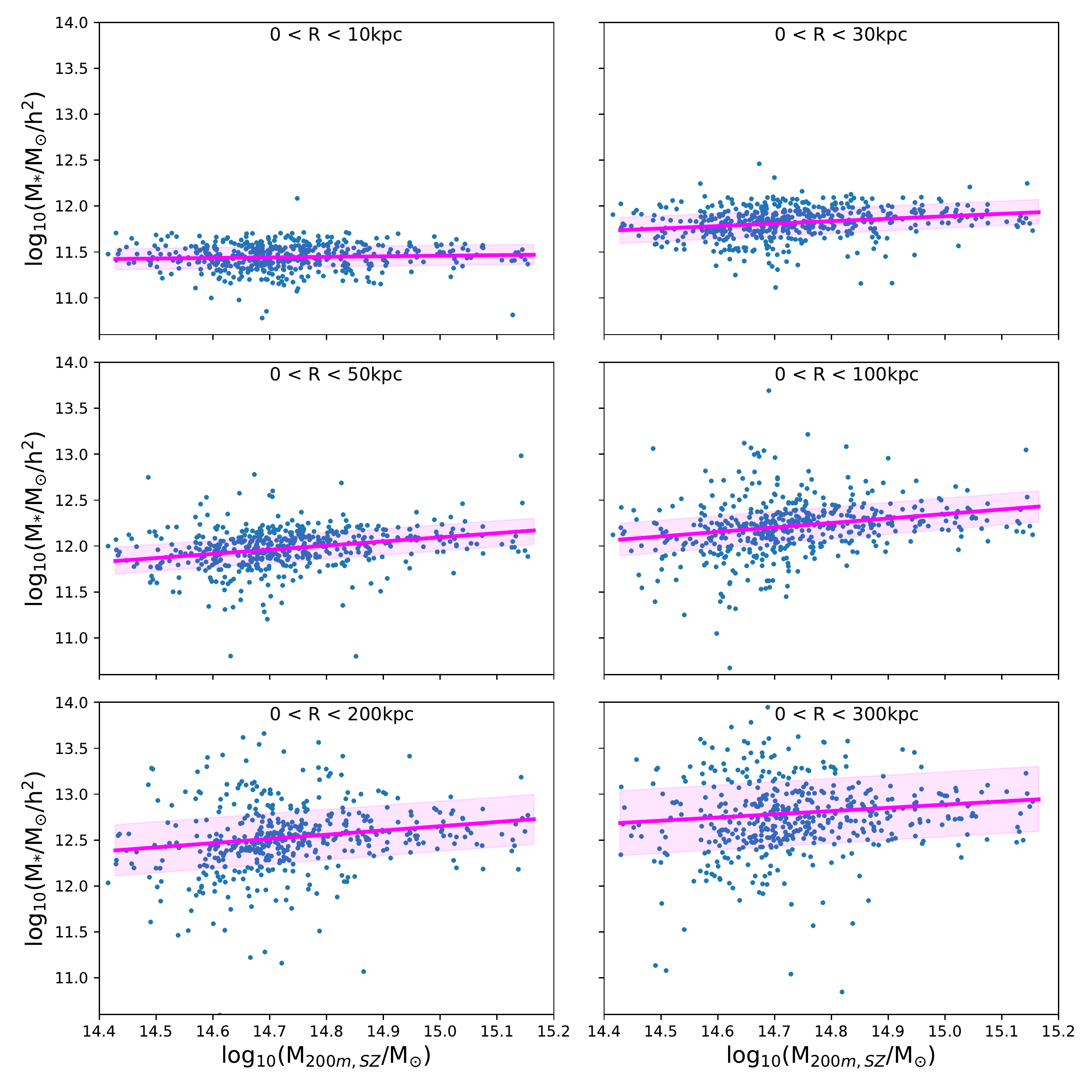}
    \caption{The total stellar mass contained within different radial regions of the BCG+ICL system, centered on the BCG, as a function of $M_{\rm 200m,SZ}$ for the DES-ACT clusters.  This version of the analysis uses the unmasked data, representative of the cluster's total stellar mass within each aperture centered on the BCG.  In magenta, we overlay the underlying distribution based on the posterior distributions provided in Table~\ref{tab:SMHM_Posteriors} for each radial bin.  These figures highlight the qualitative correlation between the total stellar mass and $M_{\rm 200m,SZ}$ and highlight that as we move to larger radii, the intrinsic scatter grows.}
    \label{fig:ICL-SMHM-nomask}
\end{figure*}

\begin{figure*}
    \centering
    \includegraphics[width=18cm]{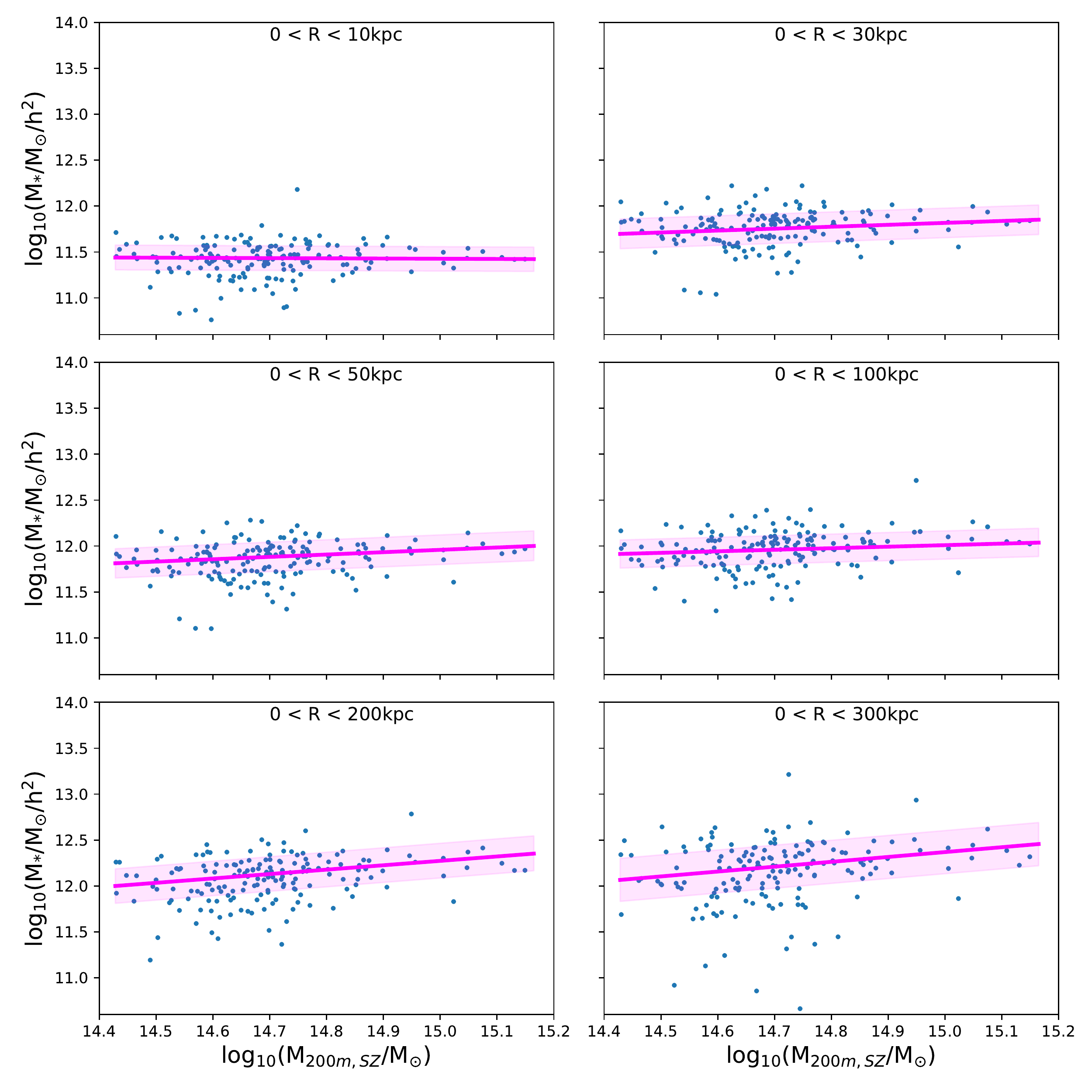}
    \caption{The BCG+ICL stellar mass contained within different physical radii as a function of $M_{\rm 200m,SZ}$ for the DES-ACT clusters.  In this version of the analysis we use the M$_{*}$+2 masking limit, given in Table~\ref{tab:limits}, which is representative of just the BCG + ICL. In magenta, we overlay the underlying distribution measured from the posterior distributions provided in Table~\ref{tab:SMHM_Posteriors} for each radial bin.   These figures highlight the correlation between the BCG+ICL stellar mass and $M_{\rm 200m,SZ}$.  Both the strength of and scatter within this correlation appear to grow as the outer radius increases.}
    \label{fig:ICL-SMHM-2mask}
\end{figure*}

In Figures~\ref{fig:ICL-SMHM-nomask} and \ref{fig:ICL-SMHM-2mask}, we illustrate the underlying stellar mass and $M_{\rm 200m,SZ}$ distributions for each of the unique annular regimes.  In Figure~\ref{fig:ICL-SMHM-nomask}, we are using the total stellar light,  while in Figure~\ref{fig:ICL-SMHM-2mask}, we show the light from the BCG+ICL based on the M$_{*}$+2 masking.  For both measurements, we use the volume complete cluster sample.  For the total light, measurements we have $\approx$ 500 clusters, while for the BCG+ICL profiles we have $\approx$ 200 clusters as a result of the strict masking.  Figures~\ref{fig:ICL-SMHM-nomask} and \ref{fig:ICL-SMHM-2mask}, visually demonstrate that the intrinsic scatter in aperture stellar mass at fixed $M_{\rm 200m,SZ}$ increases as we extend the BCG+ICL system out to larger radii.  As as result of the uncertainty associated with $M_{\rm 200m,SZ}$ and the stellar mass estimates, we take advantage of our Bayesian infrastructure, described in Section~\ref{subsec:model} to measure the parameters associated with the SMHM relation.  Figures~\ref{fig:ICL-SMHM-nomask} and ~\ref{fig:ICL-SMHM-2mask} show the median and standard deviation of a bootstrap sampling of the underlying distribution in stellar and halo mass constructed using the posterior distribution for each parameter from our Bayesian analysis.   

\subsection{Statistical Modelling of the SMHM Relation}
\label{subsec:model}
We use a nearly identical hierarchical Bayesian MCMC approach to what is described in \citet{gol19} to measure the parameters of the SMHM relation; $\alpha$, the mathematical offset; $\beta$, the slope; and $\sigma_{int}$, the intrinsic scatter in stellar mass at fixed halo mass, as given in Equation~(\ref{eq:SMHM}).  The Bayesian formalism works by convolving prior information associated with a selected model with the likelihood of the observations given that model, yielding the posterior distribution of the parameters.  To determine the posterior distributions for each SMHM parameter, our MCMC model generates values for the observed aperture stellar masses and $M_{\rm 200m,SZ}$ at each step in our likelihood analysis, which are directly compared to our observed measurements.

We model the log$\rm_{10}$ aperture stellar masses ($y$) and log$\rm_{10}(M_{\rm 200m,SZ}$/M$_{\odot})$ ($x$) as normal distributions with mean values based on our measurements. The standard deviations associated with each cluster's stellar mass and $M_{\rm 200m,SZ}$ are taken from the uncertainties on each measurement.  Since we use DES photometry, the estimate for the stellar mass uncertainty is based on the \citet{gol22} analysis, and the uncertainty for $M_{\rm 200m,SZ}$ comes from the \citet{hil20} ACT-SZ catalog.  Additionally, as done in \citet{gol19}, and \citet{gol22}, to reduce the covariance between $\alpha$ and $\beta$ we subtract off the median values of the upper and lower limits in both log$_{10}(M_{*}$/M$_{\odot})$ and log$_{10}(M_{\rm 200m,SZ}$/M$_{\odot})$, 11.5 and 14.80 respectively, from the values used in our analysis.  Lastly, as introduced in \citet{gol18}, our total uncertainty also includes an estimate of the observational uncertainty ($\sigma_{x_{0}}$ and $\sigma_{y_{0}}$) as well as a stochastic component from a beta function, $\beta(0.5,100)$, which allows us to account for the uncertainty associated with our observational error measurements.  Statistically, we treat these as free nuisance parameters, $\sigma_{x}$ and $\sigma_{y}$ in our model.

Our goal is to constrain the offset ($\alpha$), slope ($\beta$), and intrinsic scatter ($\sigma_{int}$) in the SMHM relation.  Based on the lack of detected evolution in the stellar mass of the ICL, and the results from \citet{gol22} when the magnitude gap is not incorporated, we do not study redshift evolution in this analysis.  The parameters associated with our Bayesian analysis are given in Table~\ref{tab:DESbayes}  

\begin{table*}
\centering
\caption{Bayesian Analysis Parameters for the DES-ACT Sample}
\begin{tabular}{ccc}
\hline
Symbol & Description & Prior\\
\hline
$\alpha$ & The offset of the SMHM relation & $\mathcal{U}$(-20,20) \\
$\beta$ & The high-mass power law slope & Linear Regression Prior \\ 
$\sigma_{int}$ & The uncertainty in the intrinsic stellar mass at fixed $M_{\rm 200m,SZ}$ & $\mathcal{U}(0.0,0.5)$\\ 
$y_{i}$ & The underlying distribution in stellar mass & Equation~\ref{eq:DESSMHM_redshift} \\ 
$x_{i}$ & The underlying $M_{\rm 200m,SZ}$ distribution & $\mathcal{N}$(14.23,$0.18^2$)\\
$\sigma_{y_{0i}}$ & The uncertainty between the observed stellar mass and intrinsic stellar mass distribution & 0.06 dex\\ 
$\sigma_{x_{0i}}$ & The uncertainty associated with log$_{10}$(M$_{\rm 200m,SZ}$) & $<\sigma_{x_{0i}}>$ $\approx$ 0.1  \\
\hline
\end{tabular}
\small
\\
$\mathcal{U}(a,b)$ refers to a uniform distribution where a and b are the upper and lower limits.  The linear regression prior is of the form $-1.5 \times log(1+\beta^2)$.  $\mathcal{N}(a,b)$ refers to a Normal distribution with mean and variance of a and b and that for $x_{i}$, the means and widths given in this table are example values.  The value used for $\sigma_{x_{0i}}$ is determined individually for each cluster, so the median value is provided.
\label{tab:DESbayes}
\end{table*}

As in \citet{gol18}, we model the underlying SMHM relation for each aperture as:
\begin{equation}
\label{eq:DESSMHM_redshift}
    y_{i}=\alpha + \beta*x_{i},
\end{equation} 
where $x_{i}$ and $y_{i}$ are the underlying log$_{10}$($M_{\rm 200m,SZ}$) and BCG+ICL aperture stellar masses, respectively.  We also assume a Gaussian likelihood form for $\sigma_{int}$. 

We express the entire posterior as:

\begin{equation}
p(\alpha,\beta,\sigma_{int}, x_{i},\sigma_{y_i},\sigma_{x_i} | x,y) \\ 
    \propto \underbrace{\sum_{i} P(y_{0i}|\alpha,\beta,\sigma_{y_i},\sigma_{int}, x_{i}) ~ P(x_{0i}|x_{i},\sigma_{x_i})}_{\text{likelihood}} \\ 
    \times \underbrace{p(x_i) ~  p(\sigma_{x_i}) ~ p(\sigma_{y_i}) ~ p(\alpha) ~ p(\beta) ~ p(\sigma_{int})}_{\text{priors}} 
\label{eq:DES-ACTposterior}
\end{equation}

where each $i^{th}$ cluster is a component in the summed log likelihood and the terms marked $0i$ are representative of the observed data, which as previously described is modelled as a Gaussian distribution.  

\subsection{Statistical Analysis of the BCG+ICL and Total Stellar Light SMHM relations}
\label{subsec:stat_masked_unmasked}
\begin{figure}
    \centering
    \includegraphics[width=\columnwidth]{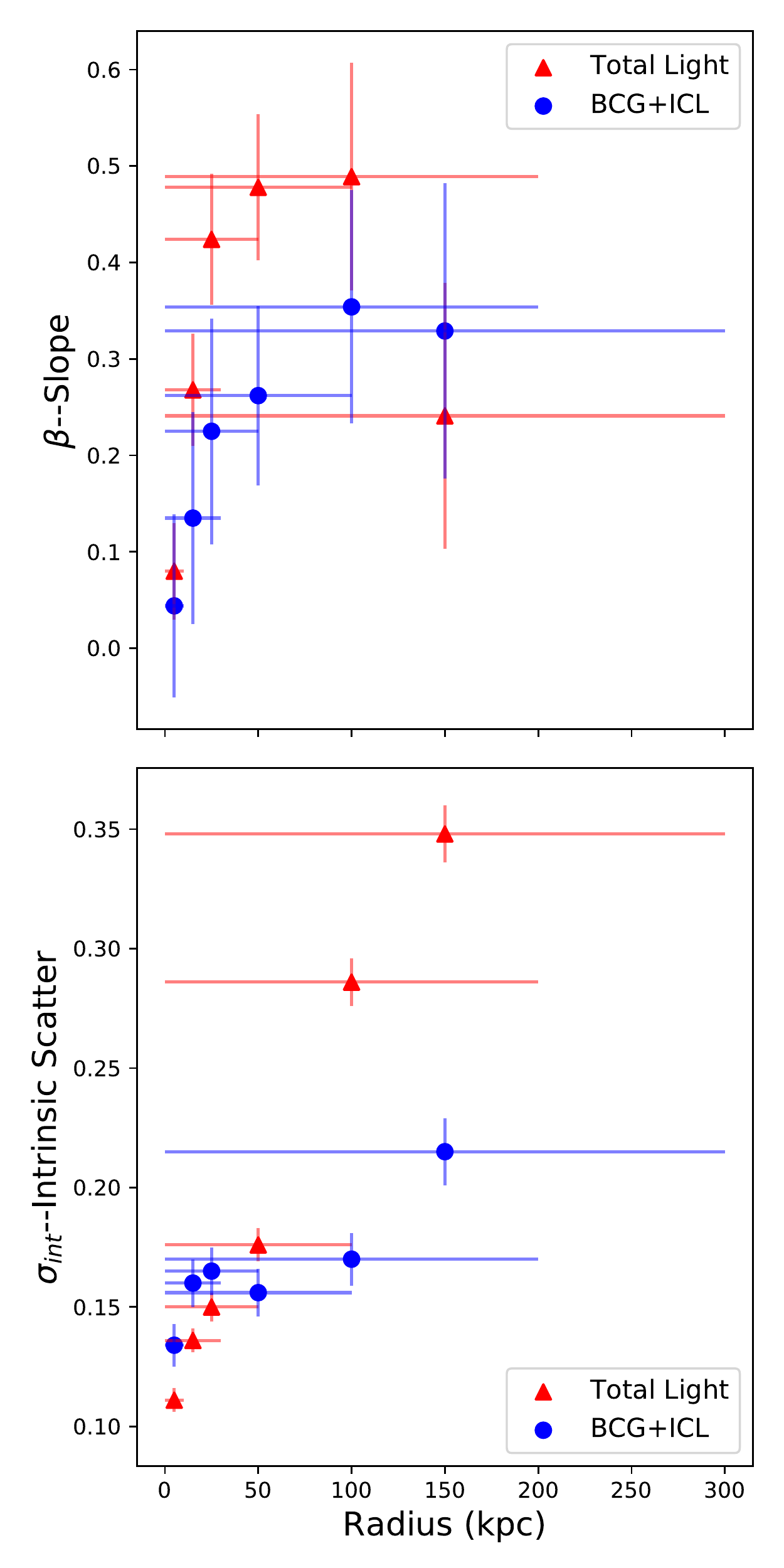}
    \caption{The slope and scatter of the SMHM relation measured for each of the 6 different aperture radial bins plotted as a function of the mean radius for both the total cluster light and BCG+ICL measurements.  The slopes and scatters are estimated using the posterior distribution of our Bayesian MCMC analysis.  The errorbars in the radius are represent the radial range over which this measurement was taken.  We measure a significantly smaller scatter and a comparable slope for the BCG+ICL measurement.}
    \label{fig:ICLslope}
\end{figure}

To better understand the impact of aperture on our SMHM relation parameters, in Figure~\ref{fig:ICLslope} we illustrate how $\beta$ and $\sigma_{int}$ vary with outer radius.  The values of the three parameters of our posterior distributions are provided in Table~\ref{tab:SMHM_Posteriors}. Additionally, although these parameters are related ($\beta$ and $\sigma_{int}$), they are not covariant in our model (the covariance between $\alpha$ and $\beta$ is reduced by median subtraction).  Additionally, in an idealised model we desire a large $\beta$ and a small $\sigma_{int}$.  So, these are the criteria we compare the SMHM relation for the total stellar light within the cluster to the version measuring the light associated within only the BCG+ICL system.  

In Figure~\ref{fig:ICLslope} we include our measurements performed for the total cluster light and the BCG+ICL system (in blue and red, respectively).  The masking process, which yields the BCG+ICL system's light, removes light associated with neighboring galaxies, therefore, it is unsurprising that the BCG+ICL measurements have a significantly lower $\sigma_{int}$ ($> 5\sigma$) at larger radii than the total stellar light versions.  The difference in slope measurements is slightly more nuanced.  At smaller radii, the slope is steeper when measured using the total stellar light of the cluster.  However, as we extend out to the largest radii, we are more heavily influenced by bright background and foreground galaxies as well as infalling satellites, which since they are unconnected to the underlying dark matter halo, may weaken the correlation and increase $\sigma_{int}$.  Based on the results shown in Figure~\ref{fig:ICLslope}, it's clear that using the total stellar mass within 100\,kpc may yield the strongest correlation.  However, because we're interested in the impact of the ICL (between 50\,kpc and 300\,kpc), the total stellar mass versions are not as useful.  At large radii, the slopes are within 1$\sigma$ of one another while $\sigma_{int}$ is significantly lower using the BCG+ICL measurement.  Therefore, going forward, we focus our analysis on the BCG+ICL system rather than further studying the SMHM relation for the total cluster light.  

For the BCG+ICL SMHM relation measurement, we see that the slope continues to increase out to radii of 200\,kpc.  However, the value measured between 0-200\,kpc and 0-300\,kpc are within 1$\sigma$ of the value measured out to 50\,kpc and 100\,kpc so, any continued evolution does not appear to be statistically significant.  Additionally, the general trend we measure agrees with those found in \citet{gol19} and \citet{mos18}.  Moreover, the asymptoting slope of $\approx$ 0.3-0.4, is in excellent agreement with the asymptoting values measured in both \citet{zhang16} and \citet{gol19}.

\subsection{The Impact of the BCG's Core on the SMHM Relation}
\label{subsec:aperture}
As a result of the two-phase formation scenario, the increase in the slope of the SMHM relation with radius results from the growth of the outskirts of the BCG.  Therefore, it's possible that this correlation may increase if the stellar mass contained within the core of the BCG+ICL system, or rather the BCG itself, is excluded, leaving only the diffuse ICL.  This idea is supported by the results from \citet{hua21}, who measure a stronger correlation between an annular stellar mass (between 50-100\,kpc) and halo mass than when the BCG's core is included. 

\begin{figure*}
    \centering
    \includegraphics[width=18cm]{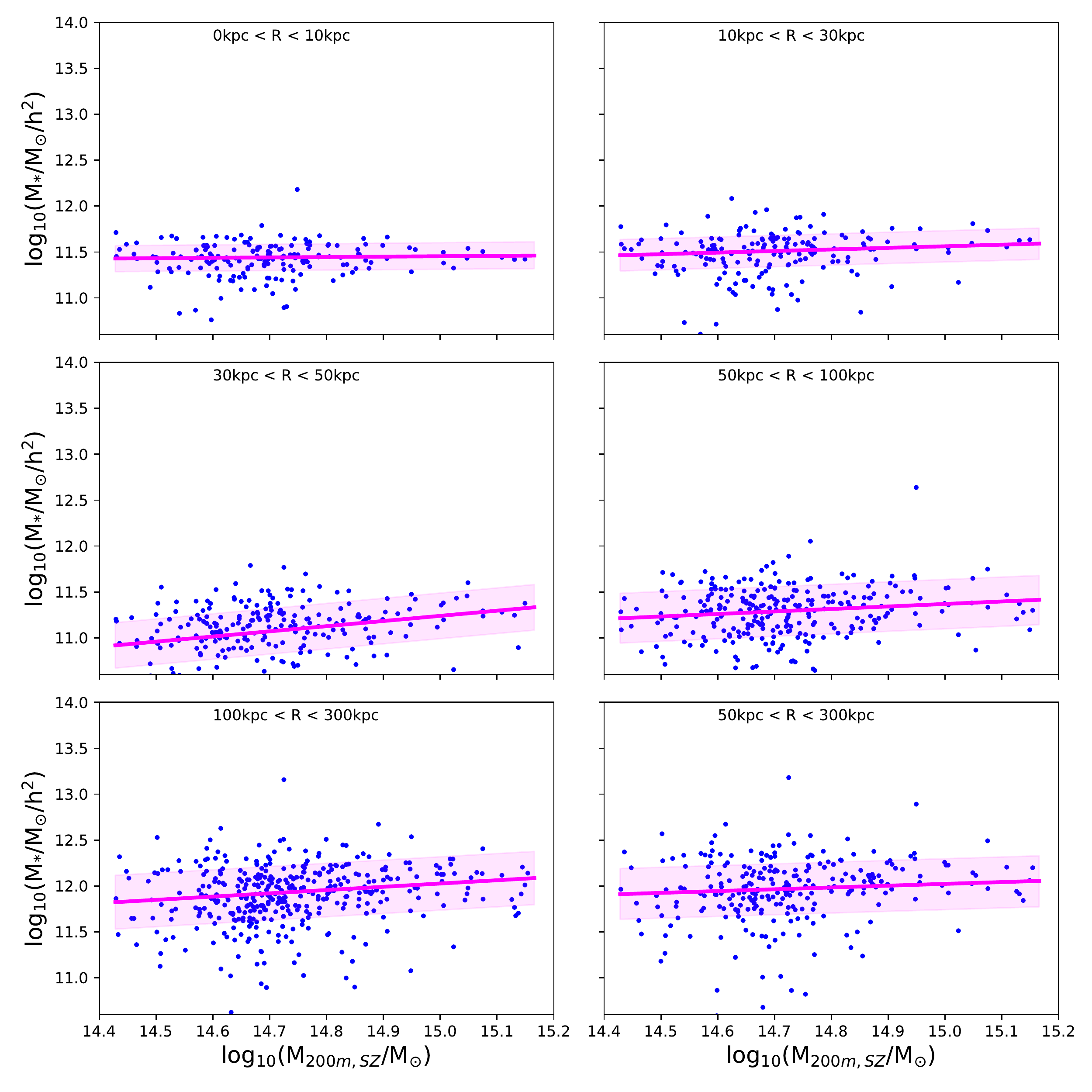}
    \caption{The stellar mass contained within different apertures of the BCG+ICL system as a function of $M_{\rm 200m,SZ}$ for the DES-ACT clusters.  In this version of the analysis we used the M$_{*}$+2 masking limit, like in Figure~\ref{fig:ICL-aperture-2mask}.  The magenta line and band are representative of the underlying distribution based on the posterior distributions provided in Table~\ref{tab:SMHM_Posteriors} for each radial bin.   This figure highlights that the slope of the SMHM relation is higher for apertures that exclude the core, but that the intrinsic scatter is much larger.}
    \label{fig:ICL-aperture-2mask}
\end{figure*}

Using our masked photometry over the radial regimes 0-10\,kpc, 10-30\,kpc, 30-50\,kpc, 50-100\,kpc, 100-300\,kpc, as well as 50-300\,kpc, which for the purpose of this paper is representative of the entire ICL (shown in Figure~\ref{fig:ICL_Mstar}), we measure the SMHM relation in discrete bins.  Additionally, for comparison to measurements from the Illustris TNG300-1 simulation \citep{pil18}, we measure the stellar mass between annular rings of 15-50\,kpc, 50-150\,kpc, and 150-300\,kpc.  In Figure~\ref{fig:ICL-aperture-2mask}, we illustrate the underlying annular stellar masses and $M_{\rm 200m,SZ}$ distributions for each of the unique annular regimes.  Because of the overlap in the radial regimes, we do not include the bins used in the TNG comparison.  For this analysis, we again use the volume complete cluster sample.  Additionally, unlike with the measurements centered at 0.0, because not every cluster has each radial measurement (i.e., some clusters lack the core measurements as a result of our masking procedure), the number of clusters in each annular bin generally increases from $\approx$~200 clusters to $\approx$~300 clusters as we move from the core to the ICL.

Figure~\ref{fig:ICL-aperture-2mask}, visually demonstrates that the intrinsic scatter in aperture stellar mass at fixed $M_{\rm 200m,SZ}$ increases as we extend our analysis out to annuli that extend into the ICL.  Moreover, it suggests that the scatter is larger than what is shown in Figure~\ref{fig:ICL-SMHM-2mask}.  Using our Bayesian infrastructure, we measure the SMHM relation parameters for each aperture.  Based on the 1D posterior distributions associated with the three parameters of the SMHM relation ($\alpha$, $\beta$, and $\sigma_{int}$) we quantify how these parameters change as we extend our analysis from the core to the ICL. The underlying correlation determined from these posterior distributions are overlayed in Figure~\ref{fig:ICL-aperture-2mask} in magenta.

\begin{figure}
    \centering
    \includegraphics[width=\columnwidth]{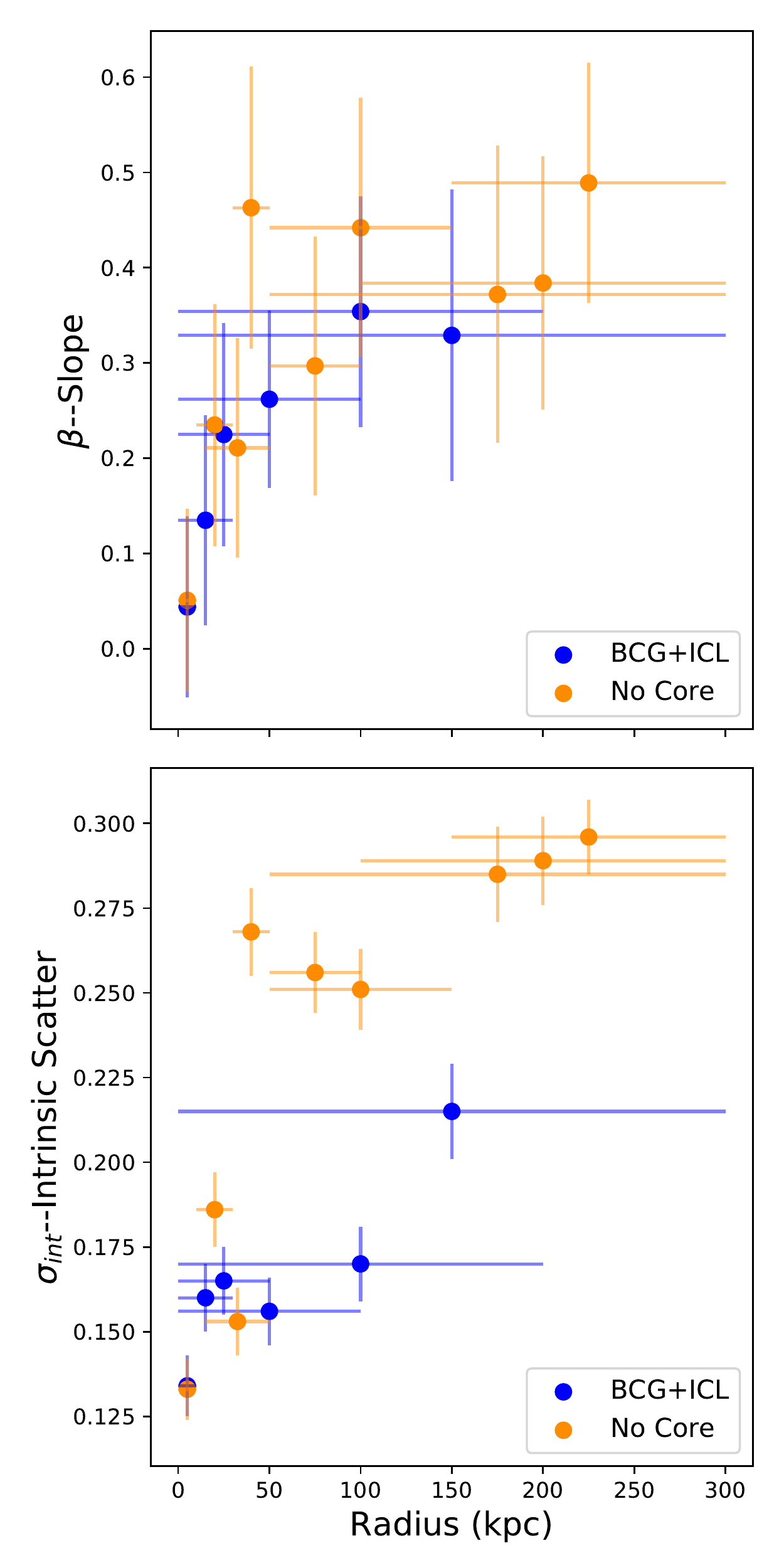}
    \caption{$\beta$ and $\sigma_{int}$ measured for each of the different radial aperture bins shown as a function of the mean radius.  We plot the versions representative of the BCG+ICL system, in blue, and the versions measured excluding the core in orange.  Here, the values are estimated as the mean and standard deviations of the posterior distributions of our Bayesian MCMC analysis.  The errorbars in the radius are representative of the radial range over which this measurement was taken.  For both versions, we find a slightly steeper slope (though within 1$\sigma$) and a higher scatter in the outer portion of the BCG+ICL system than in the core.}
    \label{fig:ICLslope_scatter_core_nocore}
\end{figure}

Moreover, in Figure~\ref{fig:ICLslope_scatter_core_nocore}, we show how the slope, $\beta$, and intrinsic scatter, $\sigma_{int}$ vary, respectively by comparing the measurements taken for the BCG+ICL system (in blue), shown in Figure~\ref{fig:ICLslope}, to the versions when the core is not included, shown in orange.  In both sets of measurements, when using only the stellar content of the core, particularly the innermost 10\,kpc, the slope of the SMHM relation is small, offering almost no correlation with $M_{\rm 200m,SZ}$.  Similarly, $\sigma_{int}$ is also small.  Such measurements support the early concept of BCG cores as standard candles \citep{san72a, san72b,gun75}.  However, when the stellar profile is extended beyond the innermost 10\,kpc (even when the core continues to be included), the SMHM relation's slope is no longer consistent with zero and thus beyond the BCG's core BCG's are no longer strong standard candles. 

In agreement with \citet{gol19}, \citet{mos18}, and \citet{hua21}, as we extend out to larger radii, the slope of the SMHM relation increases when the core is included and does so, (though within errors), until the outer aperture reaches $\approx$~100\,kpc, where the slope flattens, likely because as \citet{hua18} find, the majority of the light within the BCG+ICL system is contained within 100\,kpc. Moreover, we note that for the apertures that exclude the core (the orange points), we see a similar trend, where the slope reaches an asymptote value when the outer edge extends to 50\,kpc.

Figure~\ref{fig:ICLslope_scatter_core_nocore} illustrates that the slope of the BCG+ICL SMHM relation asymptotes at a value of $\approx$~0.30.  In comparison, when measuring the SMHM relation without light from the core, we find a slightly steeper slope that asymptotes at a value of $\approx$~0.4.  So, to first order, our results support the trend identified in \citet{hua21}; excluding the BCG's core yields a stronger correlation with $M_{\rm 200m,SZ}$ than when the core is incorporated.  As discussed in \citet{hua21}, this correlation likely results from the stellar mass in the outskirts/ICL being representative of the ex-situ stellar mass, which grows via mergers and tidal stripping, making it more representative of the recent merger growth of the underlying host halo than the core.  However, we note that the values measured when the core is excluded are within 1$\sigma$ of those measured when using entire BCG+ICL system.  Therefore, we require additional precision, through the incorporation of more clusters, to verify the results found in \citet{hua21}.

We see a similar trend for $\sigma_{int}$, such that as more light at larger radii is incorporated, we measure a larger $\sigma_{int}$.  Given that the outer envelopes are where we see divergence in BCG formation/growth, this increased scatter is unsurprising.  Additionally, we note that while the slope of the SMHM relation is shallower when the core is included, the core's ``standard candle" nature leads to a much smaller $\sigma_{int}$.  For example when comparing the radial range between 0 and 200\,kpc and the similar range of 50\,kpc to 150\,kpc the difference in $\sigma_{int}$ is approximately 0.08 dex and the values differ by more than 3.5$\sigma$.  Similarly, if we compare the radial range 0-300\,kpc to 50-300\,kpc, the $\sigma_{int}$ values differ by 0.07 dex or 2.5$\sigma$.  However, as a caveat, we note that to model the underlying halo mass distribution using the SMHM relation, as described in \citet{hua21}, the goal is both a large slope and low scatter in halo mass at fixed stellar mass.  The $\sigma_{int}$ measured in our Bayesian formalism is the intrinsic scatter in stellar mass at fixed halo mass.  As given by Equation (6) from \citet{hua21}, the scatter in halo mass at fixed stellar mass is dependent on both $\beta$ and $\sigma_{int}$, such that a higher $\beta$ and larger $\sigma_{int}$ can yield a smaller scatter in halo mass at fixed stellar mass than a lower $\beta$ and lower $\sigma_{int}$.  Therefore, in agreement we find that for some of our larger radial bins, excluding the BCG's core, does yield a tighter constraint on the halo mass, but greater precision is needed to strengthen this argument.

\section{The BCG+ICL SMHM relation in the Illustris TNG300-1 Simulation}
\label{sec:TNG}
\begin{figure}
    \centering
    \includegraphics[width=\columnwidth]{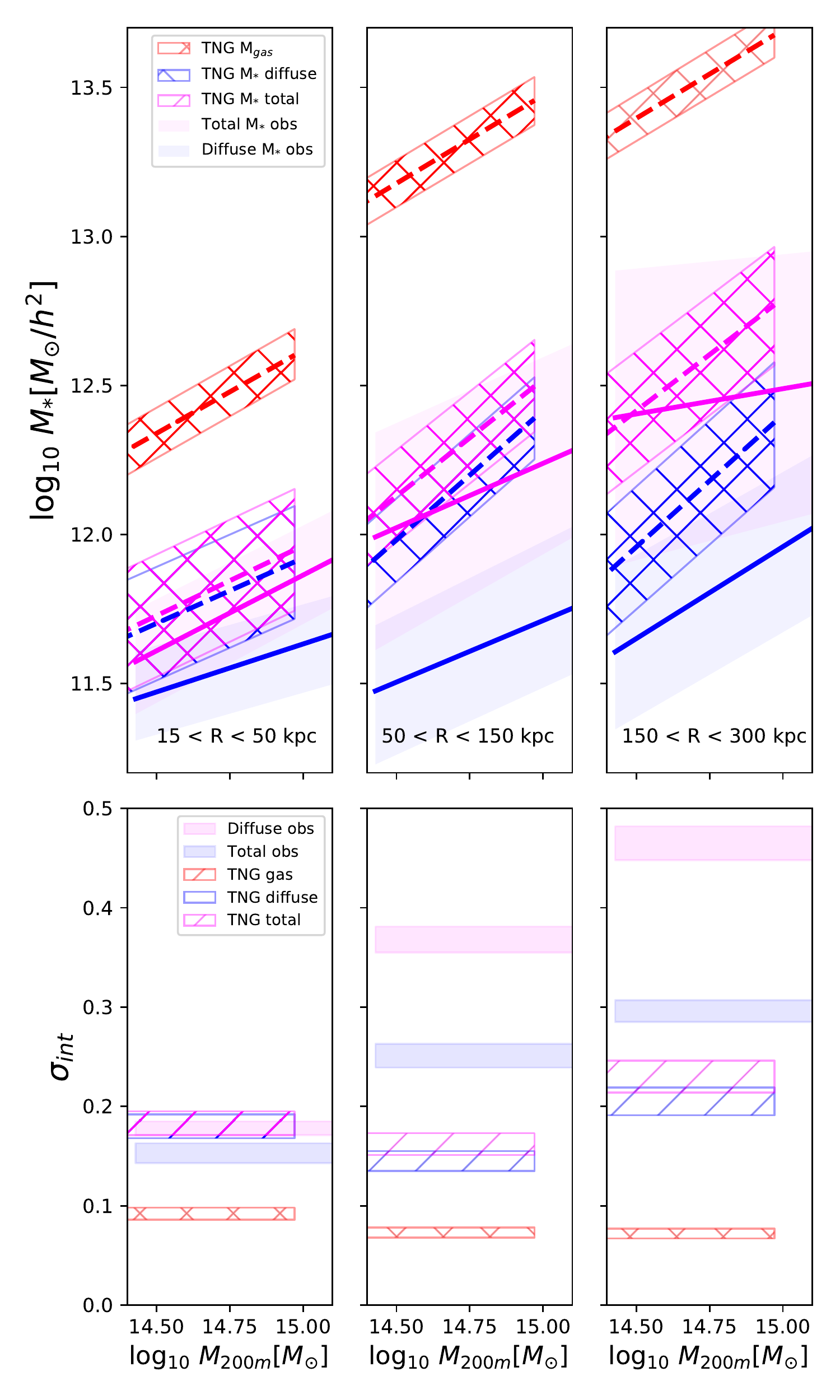}
    \caption{The posterior distribution of the SMHM relation measured in different physical apertures for the TNG data and our observational data. In the lower panels, we compare the differences in the $\sigma_{int}$ for the same mass measurements for both the observational and TNG300-1 data.}
    \label{fig:TNG}
\end{figure}

Observationally, we have shown that the correlation between the stellar mass contained within the ICL and $M_{\rm 200m,SZ}$ is stronger than what is measured for the core of the BCG+ICL system.  To further these results, we compare our measurements to measurements of high-resolution haloes from the Illustris TNG300-1\footnote{http://www.tng-project.org/} magneto-hydrodynamical cosmological simulation \citep{pil18}.  For this analysis, we use a sample of 205 clusters, with a halo mass greater than log$_{10}(M_{200m}/(M_{\odot}/h)=13.8$ (though we only show results for log$_{10}(M_{200m}/(M_{\odot}/h) > 14.4$ to match the observed measurements) in Figure~\ref{fig:TNG}, and exclude those haloes within 20Mpch$^{-1}$ of the edge of the simulation, observed at a redshift of z=0.27, 0.42, and 0.58.  Since we do not detect observational evidence of evolution, in Figure~\ref{fig:TNG}, we only display data from the z=0.58 snapshot, which is the median redshift of our observational sample.  However, the posteriors for each redshift bin are provided in Table~\ref{tab:SMHM_Posteriors}.

Following the analysis in \citet{san21}, we look at three radial ranges: 15-50\,kpc, 50-150\,kpc, and 150-300\,kpc of the BCG+ICL system.  We measure the correlation between $M_{200m}$ and the gas mass, the diffuse stellar component (the ICL), and the total stellar light using our Bayesian MCMC infrastructure.  To create a fair comparison, we subtract off the same offset as for our observational sample.  Moreover, we chose the diffuse and total light because these measurements parallel our measurements of the observational total stellar mass and stellar mass excluding the BCG core.  We note that while prior studies \citep{pil14,mon18,pil18,san21} have shown strong correlations between the diffuse light (ICL) component and halo mass in simulations, those studies do not use a similar hierarchical Bayesian MCMC approach to what we use here (described in Section~\ref{subsec:model}). 

In Figure~\ref{fig:TNG}, we show a similarly strong correlation between the TNG measurements as what exists within our observational measurements.  
In the upper row of Fig~\ref{fig:TNG}, we plot the SMHM relation for the three radial bins.  To make this figure accessible, we use the same colours for the data we compare.  Our observed sample is from clusters that are more massive than those in TNG, so we only look at the highest mass end of the TNG sample.  Moreover, although the total stellar masses are similar, since TNG300-1 is not a light cone, we cannot account for the projection effects that exist within observational data, particularly the total stellar mass estimate.

In the central region (panel 1), we see strong agreement between our measurements for the estimated stellar masses and scatter.  However, we note that the slopes of the TNG300-1 SMHM relation's are steeper.  Interestingly, these results diverge as we move to larger radii. We see that at larger radial apertures, the scatter associated with the total $M_{*}$ observed becomes significantly greater than what is measured within TNG and the amount of diffuse light though in moderate agreement appears to be greater in TNG.  The discrepancy (particularly at larger radii) in the intrinsic scatter is likely caused by the presence of foreground and background galaxies, which are included in our observational measurements, but not in the TNG snapshots.  We also note that a consistently steeper slope is found in the TNG300-1 data than in observations, in agreement with similar results found in \citet{gol22}.    

When we examine the $\sigma_{int}$ measurements for the simulation, we see that the gas mass, a well known low-scatter halo mass indicator, has the smallest scatter \citep{voi05,kra06}.  Moreover, we find that in TNG300-1, the scatter between the diffuse stellar and the total stellar components are nearly indistinguishable, hovering around 0.20 dex.  Our inner aperture, 15-50\,kpc, is in excellent agreement with the observations.  Unfortunately, as we move to larger radii, the observationally measured scatter increases dramatically to ~0.25-0.3 dex for the diffuse light and to ~0.5 dex for the total light.  However, as previously noted, the TNG300-1 measurements were taken using the 3D information of the simulation, as opposed to the 2D projected measurements for our observations, which are also impacted by foreground and background galaxies.  Therefore, it is likely that this disagreement may result from these observational uncertainties.  

Though not directly observed from Figure~\ref{fig:TNG}, we comment on how the slope and scatter vary with radius for the TNG300-1 data.  For the diffuse light (and total light), we see that as we move to apertures that extend to larger radii, the slope of the SMHM relation significantly increases.  We also see that the scatter similarly increases, though much less than in observations.  The similarity between these trends to what we detect in our observed data highlights that the outskirts of the BCG+ICL system are more strongly correlated with the halo mass of the cluster than the core of the BCG.  Howeover, because this slope is significantly steeper, this correlation is even stronger in the TNG300-1 universe than what we observe using the DES-ACT data.  This combination of a similarity and discrepancy suggests that growth of the ICL is similar to what is observed in the real universe, but that the underlying correlation between these two parameters is much stronger in the underlying models of the TNG300-1 simulation.

\begin{table*}{cccccc}
\caption{Posterior Distribution Results}
\centering
\begin{tabular}{cccccc}
\hline
Data & Inner Radius & Outer Radius & $\alpha$ & $\beta$ & $\sigma_{int}$ \\
\hline
Total & 0 & 10 &$-0.040 \pm 0.007$ & $0.074 \pm 0.048$ & $0.111 \pm 0.004$  \\
Total & 10 & 30 &$0.105 \pm 0.010$ & $0.442 \pm 0.077$ & $0.185 \pm 0.007$  \\
Total & 15 & 50 &$0.270 \pm 0.010$ & $0.526 \pm 0.081$ & $0.178 \pm 0.007$  \\
Total & 30 & 50 &$-0.049 \pm 0.014$ & $0.772 \pm 0.112$ & $0.259 \pm 0.010$  \\
Total & 50 & 100 &$0.340 \pm 0.021$ & $0.572 \pm 0.154$ & $0.384 \pm 0.013$  \\
Total  & 50 & 150 &$0.680 \pm 0.020$ & $0.562 \pm 0.150$ & $0.368 \pm 0.013$  \\
Total  & 100 & 300 &$1.086 \pm 0.028$ & $0.312 \pm 0.191$ & $0.491 \pm 0.017$  \\
Total & 50 & 300 &$1.176 \pm 0.023$ & $0.299 \pm 0.171$ & $0.439 \pm 0.015$  \\ 
Total & 150 & 300 &$0.952 \pm 0.025$ & $0.193 \pm 0.186$ & $0.465 \pm 0.017$  \\ \hline
BCG+ICL & 0 & 10 &$-0.059 \pm 0.014$ & $0.051 \pm 0.096$ & $0.133 \pm 0.009$  \\
No Core & 10 & 30 &$0.023 \pm 0.019$ & $0.235 \pm 0.127$ & $0.186 \pm 0.011$  \\
No Core & 15 & 50 &$0.088 \pm 0.017$ & $0.211 \pm 0.115$ & $0.153 \pm 0.010$  \\
No Core & 30 & 50 &$-0.366 \pm 0.023$ & $0.463 \pm 0.148$ & $0.268 \pm 0.013$  \\
No Core & 50 & 100 &$-0.206 \pm 0.019$ & $0.297 \pm 0.136$ & $0.256 \pm 0.012$  \\
No Core & 50 & 150 &$0.101 \pm 0.019$ & $0.442 \pm 0.136$ & $0.251 \pm 0.012$  \\
No Core & 100 & 300 &$0.417 \pm 0.019$ & $0.384 \pm 0.133$ & $0.289 \pm 0.013$  \\
No Core & 50 & 300 &$0.507 \pm 0.022$ & $0.372 \pm 0.156$ & $0.285 \pm 0.014$  \\ 
No Core & 150 & 300 &$0.317 \pm 0.017$ & $0.489 \pm 0.126$ & $0.296 \pm 0.011$  \\ \hline
Total & 0 & 10 &$-0.040 \pm 0.007$ & $0.080 \pm 0.050$ & $0.111 \pm 0.005$  \\
Total & 0 & 30 &$0.343 \pm 0.008$ & $0.268 \pm 0.058$ & $0.136 \pm 0.005$  \\
Total & 0 & 50 &$0.507 \pm 0.009$ & $0.424 \pm 0.068$ & $0.150 \pm 0.006$  \\
Total & 0 & 100 &$0.736 \pm 0.010$ & $0.478 \pm 0.076$ & $0.176 \pm 0.007$  \\
Total & 0 & 200 &$1.053 \pm 0.016$ & $0.489 \pm 0.118$ & $0.286 \pm 0.010$  \\
Total & 0 & 300 &$1.280 \pm 0.019$ & $0.241 \pm 0.138$ & $0.348 \pm 0.012$  \\ \hline
BCG+ICL & 0 & 10 &$-0.058 \pm 0.014$ & $0.044 \pm 0.095$ & $0.134 \pm 0.009$  \\
BCG+ICL & 0 & 30 &$0.289 \pm 0.017$ & $0.135 \pm 0.110$ & $0.160 \pm 0.010$  \\
BCG+ICL & 0 & 50 &$0.388 \pm 0.018$ & $0.225 \pm 0.110$ & $0.165 \pm 0.010$  \\
BCG+ICL & 0 & 100 &$0.496 \pm 0.014$ & $0.262 \pm 0.093$ & $0.156 \pm 0.010$  \\
BCG+ICL & 0 & 200 &$0.642 \pm 0.018$ & $0.354 \pm 0.121$ & $0.170 \pm 0.011$  \\
BCG+ICL & 0 & 300 &$0.749 \pm 0.023$ & $0.329 \pm 0.153$ & $0.215 \pm 0.014$  \\ \hline
TNG Gas SN79 & 15 & 50 &$0.917 \pm 0.032$ & $0.707 \pm 0.034$ & $0.110 \pm 0.005$  \\
TNG Gas SN79 & 50 & 150 &$1.736 \pm 0.027$ & $0.700 \pm 0.028$ & $0.092 \pm 0.005$  \\
TNG Gas SN79 & 150 & 300 &$2.013 \pm 0.023$ & $0.641 \pm 0.023$ & $0.077 \pm 0.004$  \\ 
TNG Gas SN71 & 15 & 50 &$0.884 \pm 0.035$ & $0.686 \pm 0.036$ & $0.097 \pm 0.006$  \\
TNG Gas SN71 & 50 & 150 &$1.720 \pm 0.030$ & $0.692 \pm 0.031$ & $0.082 \pm 0.005$  \\
TNG Gas SN71 & 150 & 300 &$2.003 \pm 0.026$ & $0.627 \pm 0.026$ & $0.070 \pm 0.004$  \\ 
TNG Gas SN64 & 15 & 50 &$0.981 \pm 0.040$ & $0.612 \pm 0.041$ & $0.092 \pm 0.006$  \\
TNG Gas SN64 & 50 & 150 &$1.800 \pm 0.032$ & $0.629 \pm 0.033$ & $0.073 \pm 0.005$  \\
TNG Gas SN64 & 150 & 300 &$2.061 \pm 0.031$ & $0.568 \pm 0.032$ & $0.072 \pm 0.005$  \\ \hline
TNG Diffuse SN79 & 15 & 50 &$0.505 \pm 0.048$ & $0.661 \pm 0.049$ & $0.160 \pm 0.008$  \\
TNG Diffuse SN79 & 50 & 150 &$0.648 \pm 0.046$ & $0.800 \pm 0.047$ & $0.158 \pm 0.008$  \\
TNG Diffuse SN79 & 150 & 300 &$0.668 \pm 0.063$ & $1.024 \pm 0.065$ & $0.212 \pm 0.011$  \\ 
TNG Diffuse SN71 & 15 & 50 &$0.473 \pm 0.061$ & $0.624 \pm 0.063$ & $0.170 \pm 0.010$  \\
TNG Diffuse SN71 & 50 & 150 &$0.630 \pm 0.057$ & $0.787 \pm 0.059$ & $0.160 \pm 0.009$  \\
TNG Diffuse SN71 & 150 & 300 &$0.631 \pm 0.078$ & $0.943 \pm 0.081$ & $0.216 \pm 0.013$  \\ 
TNG Diffuse SN64 & 15 & 50 &$0.459 \pm 0.079$ & $0.582 \pm 0.080$ & $0.180 \pm 0.012$  \\
TNG Diffuse SN64 & 50 & 150 &$0.704 \pm 0.063$ & $0.843 \pm 0.065$ & $0.145 \pm 0.010$  \\
TNG Diffuse SN64 & 150 & 300 &$0.710 \pm 0.089$ & $0.993 \pm 0.091$ & $0.205 \pm 0.014$  \\ \hline
TNG Total SN79 & 15 & 50 &$0.545 \pm 0.047$ & $0.658 \pm 0.049$ & $0.162 \pm 0.008$  \\
TNG Total SN79 & 50 & 150 &$0.758 \pm 0.050$ & $0.735 \pm 0.052$ & $0.171 \pm 0.009$  \\
TNG Total SN79 & 150 & 300 &$0.842 \pm 0.066$ & $0.861 \pm 0.068$ & $0.224 \pm 0.011$  \\ 
TNG Total SN71 & 15 & 50 &$0.524 \pm 0.062$ & $0.628 \pm 0.064$ & $0.171 \pm 0.010$  \\
TNG Total SN71 & 50 & 150 &$0.716 \pm 0.067$ & $0.694 \pm 0.069$ & $0.187 \pm 0.011$  \\
TNG Total SN71 & 150 & 300 &$0.833 \pm 0.083$ & $0.811 \pm 0.085$ & $0.227 \pm 0.013$  \\ 
TNG Total SN64 & 15 & 50 &$0.491 \pm 0.078$ & $0.549 \pm 0.080$ & $0.183 \pm 0.012$  \\
TNG Total SN64 & 50 & 150 &$0.819 \pm 0.069$ & $0.765 \pm 0.070$ & $0.162 \pm 0.011$  \\
TNG Total SN64 & 150 & 300 &$0.867 \pm 0.100$ & $0.755 \pm 0.101$ & $0.230 \pm 0.016$  \\ \hline
\\
\end{tabular}
\small
\\
Although we don't include the unmasked measurements when the core is not included in Figure~\ref{fig:ICLslope_scatter_core_nocore}, we include these measurements here in the first 9 rows of this table.
\label{tab:SMHM_Posteriors}
\end{table*}

\section{The Colour of the ICL}
\label{sec:colour}
As discussed in Section~\ref{sec:intro}, the ICL is thought to grow as a result of either major/minor mergers and/or the stripping of satellite galaxies.  Colour provides a key insight into the stellar population and growth of the BCG+ICL system.  Therefore, it is unsurprising that prior observations and simulations have measured BCG + ICL colours \citep[e.g.,][]{dem15,mor17,dem18,con19,zha18,che22}.  If the ICL growth is dominated by galaxy-galaxy mergers then no colour gradient would be expected since such mergers would mix stellar populations, flattening any pre-existing colour gradient \citep{con19}, as found in some recent model dependent observational results \citep{bur15, gro17}.  However, some observations detect a negative colour gradient, such that the colour in the ICL is bluer than the colour in the BCG's core \citep{dem15,dem18,mor17, hua18, zha18, che22}, which would result if instead the ICL forms as a result of the tidal stripping of metal poor satellite galaxies \citep{mon18,con19} or minor mergers with those same galaxies \citep{hua18}.  Although the purely observational analyses favor the presence of a gradient, there exists some uncertainty about which of these methods dominates ICL growth.

\begin{figure*}
    \centering
    \includegraphics[width=18cm]{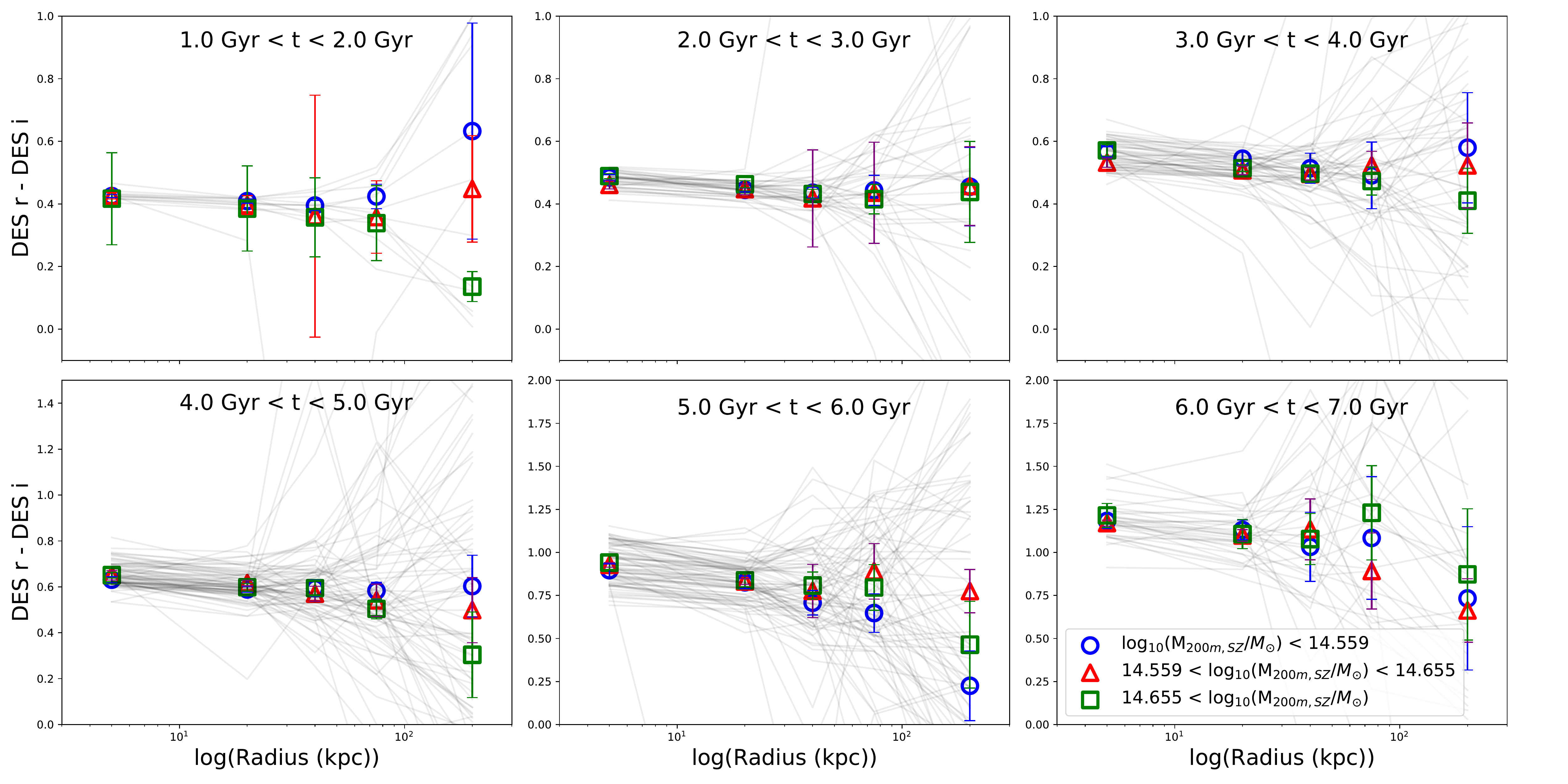}
    \caption{For each range in lookback time, we plot the colour as a function of radius (in gray).  The median colour as a function of radius for the three bins in $M_{\rm 200m,SZ}$ is overplotted.}
    \label{fig:colour_bin_Mhalo}
\end{figure*}

In Figure~\ref{fig:colour_bin_Mhalo}, using the M$_{*}$+2 masked data, described in Section~\ref{subsec:ICL_measurement}, we examine how the colour of the BCG+ICL system changes with both radius and lookback time.  The data used in this portion of the analysis come from a different subsample than what was described in Section~\ref{sec:measurements}.  Instead of using the volume complete sample, here we utilise a sample of 287 clusters that we measure both an r-band and i-band magnitude for within each of the 5 distinct radial bins (0-10\,kpc, 10-30\,kpc, 30-50\,kpc, 50-100\,kpc, and 100-300\,kpc).  As previously noted, for certain clusters, we are unable to measure a central magnitude as a result of masking due to a crowded central region, which results in the exclusion of a large portion of the DES-ACT clusters.  In Figure~\ref{fig:colour_bin_Mhalo}, we plot the colour gradient (along with the bootstrapped medians) for 6 bins divided by lookback time.  Figure~\ref{fig:colour_bin_Mhalo} shows a moderately negative colour gradient, such that the colours are bluer at large radii than in the BCG's core.  This detection follows from the expectation that the stars found in the ICL are likely younger than those found in the canonical ``red and dead'' core of the BCG. Therefore, our ICL colours agree with the results that support that the ICL's formation is dominated by the tidal stripping of satellite galaxies \citep{dem15,dem18,mor17,mon18}.  Moreover, the difference between the median colour of the BCG's core and the median colour of the ICL is small, on the order of $\approx$0.1-0.2 magnitudes, which is in agreement with the modest colour gradient that \citet{con19} measures in their ICL simulations. 

\begin{figure}
    \centering
    \includegraphics[width=\columnwidth]{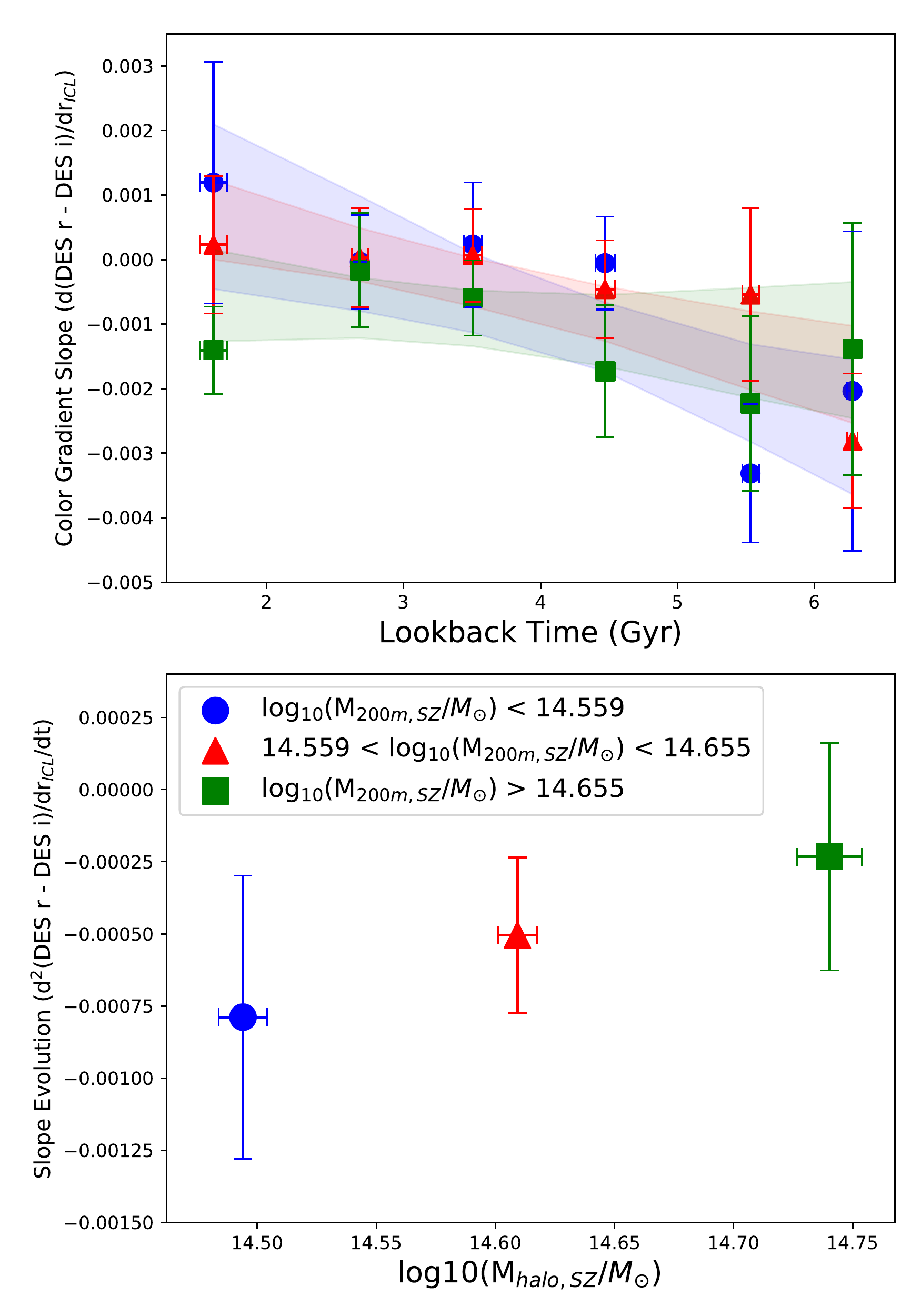}
    \caption{For the samples shown in Figure~\ref{fig:colour_bin_Mhalo}, in the upper plot, we show how the colour gradient in each $M_{\rm 200m,SZ}$ bin changes with lookback time.  In the lower figure, we show how the change in the slope of the colour gradient as a function of lookback time evolves with $M_{\rm 200m,SZ}$.}
    \label{fig:colour_bin_SZ_slope}
\end{figure}

Although an underlying trend is observed within Figure~\ref{fig:colour_bin_Mhalo}, a large amount of scatter exists within our colour measurements (shown by the noisy gray lines).  To determine if this noise is statistical or caused by a latent parameter, in Figure~\ref{fig:colour_bin_Mhalo}, we show whether any trends exist between the colour and $M_{\rm 200m,SZ}$.  
As represented by the points shown in Figure~\ref{fig:colour_bin_Mhalo}, we divide our sample into 3 $M_{\rm 200m,SZ}$ bins with the same number of clusters.  For the sample as a whole, we apply a bootstrapping technique and randomly select 287 clusters from our sample 1000 times.  For each bootstrap sample, we calculate the mean colour and standard deviation based on the draws from all 1000 bootstrapped samples, which we illustrate in Figure~\ref{fig:colour_bin_Mhalo}.  Therefore, each point is representative of the median and standard deviation of the bootstrapped median values, not the black distribution (the gray lines are what the bootstrap is drawn from).  Moreover, We note that for each bootstrap sample, the divisions in $M_{\rm 200m,SZ}$ are based on the draw of clusters for that bootstrap resampling, as shown by the uncertainty in the $M_{\rm 200m,SZ}$ measurement in Figure~\ref{fig:colour_bin_SZ_slope}.  Moreover, while the radial bins and lookback time bins are fixed, we don't require the lookback time distribution to be identical in each bootstrap sampling; as a result, there is uncertainty in the lookback time measurement in Figure~\ref{fig:colour_bin_SZ_slope}.  For each bin, we determine the median colour measured within each of the 5 radial regimes.  Based on this analysis, there is no identifiable trend between $M_{\rm 200m,SZ}$ and the colour of each BCG+ICL system.  However, it's possible that we are examining too small a regime in $M_{\rm 200m,SZ}$ parameter space to detect any such differences.    

Using the median colours we estimate the slope of the colour gradient at each of the different $M_{\rm 200m,SZ}$ bins for each bin in lookback time between 1 and 7 Gyrs using a $\chi^{2}$ optimisation fitting. 
In the upper panel of Figure~\ref{fig:colour_bin_SZ_slope}, we illustrate how the slope of the colour gradient changes across the radial extent of the BCG+ICL system as a function of lookback time for the three different $M_{\rm 200m,SZ}$ bins.  Mathematically, this is \(\frac{d(R-I)}{dr_{ICL}}\), where a value of zero means that the colour is constant across the entire radial regime.  A negative value means that the colour is bluer in the outskirts of the ICL and a positive value that it is redder in the ICL than in the BCG.  In this figure, the error bars represent the 1$\sigma$ values for the colour gradient slope based on our previously described bootstrap measurements.  The shaded region is representative of the 1$\sigma$ distribution of the fit between the colour gradient slope and lookback time for each of the $M_{\rm 200m,SZ}$ bins.  We note that the colour gradients are either consistent with zero or negative and that the slopes appear to be slightly more negative at earlier lookback times.  While the quantitative results support that the ICL grows through tidal stripping (because a negative, blue, gradient is detected), they also suggest that other mechanisms of growth may occur.  Since these are the median colour gradients, it's possible that both tidal stripping and mergers occur and that stripping may dominate such growth at earlier times.  

In the lower panel of Figure~\ref{fig:colour_bin_SZ_slope} we illustrate how the slope of the colour gradient evolves with lookback time as a function of $M_{\rm 200m,SZ}$.  Mathematically, this is \(\frac{d^2(R-I)}{dr_{ICL}dt}\). The error bars in both lookback time and the slope evolution are representative of the 1$\sigma$ distribution from the bootstrap measurements.  In this context, a value of zero for \(\frac{d^2(R-I)}{dr_{ICL}dt}\) would mean that the colour gradient does not change over lookback time.  A positive value is representative of both first derivatives, with respect to time and radial extent, having the same sign, for example, that the colour gradient becomes bluer over both radial extent and time.  In contrast, a negative value would occur when both first derivatives have different signs, for example the colour gradient becomes bluer over radial extent, but redder over time.  

The upper panel of Figure~\ref{fig:colour_bin_SZ_slope} shows that as M$_{\rm 200m,SZ}$ increases, the slope of the gradient across cosmic time becomes slightly more negative.  Interestingly, the lower panel shows that the change in colour with respect to both radius and time, \(\frac{d^2(R-I)}{dr_{ICL}dt}\), becomes closer to 0 as M$_{\rm 200m,SZ}$ increases.  While this trend is not statistically significant, we note that to first order, such a trend suggests that over cosmic time and radial extent, the colour gradient, out to a radius of $\approx$ 300\,kpc, is not changing for the most massive clusters, and consistently shows a negative blue colour gradient across cosmic time.  In contrast, for the lowest mass haloes, we detect a change in the colour gradient, such that the second derivative of the colour is more than 1$\sigma$ below 0.0. As such, the colour gradient is negative, with bluer colours at larger radial extents and larger lookback times (higher redshifts).  If, as expected, the colour gradient results from the tidal stripping and accretion of stars from less massive satellites, this result would suggest that the ICL's stellar content is changing over this redshift range.  However, as shown in Figure~\ref{fig:ICL_Mstar}, we don't find the stellar content of the ICL to be increasing over cosmic time.  Therefore, we limit our interpretation to just an ICL, truncated at 300\,kpc, that is not growing in stellar mass.  In this scenario, it is likely that for the more massive haloes, there has been no recent accretion within 300\,kpc.  In contrast, for the lower mass clusters we measure a \(\frac{d^2(R-I)}{dr_{ICL}dt}\), such that the value is negative for higher redshifts, and positive for lower redshifts, which may suggest that those lower mass clusters experienced recent enough tidal stripping that we only see the presence of bluer stars at higher lookback times.  Alternatively, our results may suggest that for the most massive clusters (in green), tidal stripping is the dominant growth mechanism, whereas for the less massive haloes, the change in the gradient may be caused by mergers, which wash out the gradient, but do not increase the stellar mass within the radial regime of 50\,kpc to 300\,kpc.  Of additional interest, Figure~\ref{fig:colour_bin_SZ_slope} shows that around a lookback time of $\approx$~4Gyr for all clusters, the slope of the colour gradients seems to increase slightly, becoming more consistent with 0.0, which would suggest that in the last 4 Gys, regardless of $M_{\rm 200m,SZ}$ the colour of the ICL has been relatively fixed, serving as further evidence for a lack of recent growth in the ICL between 50\,kpc and 300\,kpc.  As a caveat, we note that these results are less significant when DESY3 richness is used to bin the clusters in lieu of $M_{\rm 200m,SZ}$, which may result from the larger scatter associated with the scaling relation estimated halo masses.

\section{Conclusion}
\label{sec:conclusion}
In this analysis, we focus on characterising the properties associated with the stellar content of the BCG+ICL across redshift space using the DES-ACT sample observed over the redshift range $0.2 < z < 0.8$.  Since it is difficult to disentangle the outskirts of the BCG from the underlying ICL, we do not attempt to distinguish the two and instead treat the inner 50\,kpc as being associated with the BCG and the region between 50\,kpc and 300\,kpc as being the ICL.  In this analysis, we study the stellar content of the BCG+ICL system in three ways, the total stellar mass, the SMHM relation, and the BCG+ICL colour.  We summarise the primary results of our analysis as follows.  
\begin{itemize}
    \item We are unable to detect noticeable evolution in the stellar mass contained within the ICL (between 50-300\,kpc) when we bin the data evenly in redshift space over the redshift range $0.2 < z < 0.8$.
    \item We find that $\beta$ and $\sigma_{int}$, the slope and intrinsic scatter for the SMHM relation, increase as we incorporate BCG+ICL stellar mass information at larger radii.  This trend agrees with the results from \citet{gol19}, but extends beyond the 100\,kpc studied in that analysis.
    \item By comparing the posteriors for the SMHM relation when the BCG's core is and is not included, we find that when the core is excluded, we measure a slightly steeper slope.  This stronger correlation between the stellar outskirts and $M_{\rm 200m,SZ}$ supports that the ICL, like the BCG, grows via the two-phase formation scenario (either through tidal stripping of satellites or mergers).  Additionally, although we find a stronger correlation with $M_{\rm 200m,SZ}$, the associated intrinsic scatter is much larger when the BCG is not included.  This trend results from the absence of the stellar information from the BCG, which has a smaller intrinsic scatter. 
    \item By comparing our observational results to those from Illustris TNG300-1, we find a similar amount of diffuse light within each radial bin.  However, both $\beta$ and $\sigma_{int}$ differ.  $\sigma_{int}$ is likely larger in our observational data due to projection effects.  However, the steeper $\beta$ may highlight that the underlying correlation between the BCG+ICL system and host halo is much tighter in the simulated universe than what we observe, which would suggest further areas for improvement in simulated clusters.  
    \item We detect a modest colour gradient in the ICL in each of the lookback time bins, such that the colour of the ICL on average becomes bluer as we move out to larger radii in agreement with prior observational results \citep[e.g.,][]{dem15,dem18,mor17,mon18} that support the tidal stripping of satellites as the primary mechanism for ICL growth.  However, we measure a rather modest non-zero gradient, therefore it is possible that both processes occur and are responsible for ICL growth.    
    \item When we measure the slope of the colour gradient, averaged over bins in $M_{\rm 200m,SZ}$, we find that the change in this slope with lookback time is more than 1$\sigma$ from zero for the lowest $M_{\rm 200m,SZ}$ bin, while statistically equivalent to zero for the highest $M_{\rm 200m,SZ}$ bin.  This may suggest that the clusters within the lowest $M_{\rm 200m,SZ}$ bins show evidence of more recent merger accretion, the presence of bluer stars at the earlier lookback times, while the clusters in the higher $M_{\rm 200m,SZ}$ bins show no such evidence of recent growth or that the lowest mass haloes experience recent growth from mergers.  
\end{itemize}   
    
Going forward, there are many avenues of research we aim to explore to improve our observational understanding and characterisation of the ICL.  In future analyses, we plan to further examine the trend between the colour, halo mass, and lookback time to see if this trend persists at lower halo masses, potentially using the full DESY6 redMaPPer sample, which extends to lower halo masses than the DES-ACT overlap.  Additionally, using the DES-redMaPPer catalogs, we aim to build on this work and work presented in \citet{gol22} to determine whether any trends exist between the properties of the ICL and the magnitude gap, including determining whether the trends found in \citet{gol22} persist when the ICL is incorporated.  Moreover, we plan to investigate whether clusters characterised by large ICL fractions correlate with those characterised by large magnitude gaps, since both properties are used as observational tracers of relaxed clusters. 
\\
\\
We note that this paper has gone through internal review by the DES collaboration.  JGM would like to thank Xiaokai Chen and Ying Zu for useful discussions about the properties and characteristics of the ICL.  JGM also acknowledges the support from the National Key Basic Research and Development Program of China (No. 2018YFA0404504) and the National Science Foundation of China (No. 11873038, 11890692, and 12173024 ).

Funding for the DES Projects has been provided by the U.S. Department of Energy, the U.S. National Science Foundation, the Ministry of Science and Education of Spain, the Science and Technology Facilities Council of the United Kingdom, the Higher Education Funding Council for England, the National Center for Supercomputing Applications at the University of Illinois at Urbana-Champaign, the Kavli Institute of Cosmological Physics at the University of Chicago, the Center for Cosmology and Astro-Particle Physics at the Ohio State University, the Mitchell Institute for Fundamental Physics and Astronomy at Texas A\&M University, Financiadora de Estudos e Projetos, Funda{\c c}{\~a}o Carlos Chagas Filho de Amparo {\`a} Pesquisa do Estado do Rio de Janeiro, Conselho Nacional de Desenvolvimento Cient{\'i}fico e Tecnol{\'o}gico and the Minist{\'e}rio da Ci{\^e}ncia, Tecnologia e Inova{\c c}{\~a}o, the Deutsche Forschungsgemeinschaft and the Collaborating Institutions in the Dark Energy Survey. 

The Collaborating Institutions are Argonne National Laboratory, the University of California at Santa Cruz, the University of Cambridge, Centro de Investigaciones Energ{\'e}ticas, Medioambientales y Tecnol{\'o}gicas-Madrid, the University of Chicago, University College London, the DES-Brazil Consortium, the University of Edinburgh, the Eidgen{\"o}ssische Technische Hochschule (ETH) Z{\"u}rich, Fermi National Accelerator Laboratory, the University of Illinois at Urbana-Champaign, the Institut de Ci{\`e}ncies de l'Espai (IEEC/CSIC), the Institut de F{\'i}sica d'Altes Energies, Lawrence Berkeley National Laboratory, the Ludwig-Maximilians Universit{\"a}t M{\"u}nchen and the associated Excellence Cluster Universe, the University of Michigan, NSF's NOIRLab, the University of Nottingham, The Ohio State University, the University of Pennsylvania, the University of Portsmouth, SLAC National Accelerator Laboratory, Stanford University, the University of Sussex, Texas A\&M University, and the OzDES Membership Consortium.

Based in part on observations at Cerro Tololo Inter-American Observatory at NSF's NOIRLab (NOIRLab Prop. ID 2012B-0001; PI: J. Frieman), which is managed by the Association of Universities for Research in Astronomy (AURA) under a cooperative agreement with the National Science Foundation.

The DES data management system is supported by the National Science Foundation under Grant Numbers AST-1138766 and AST-1536171. The DES participants from Spanish institutions are partially supported by MICINN under grants ESP2017-89838, PGC2018-094773, PGC2018-102021, SEV-2016-0588, SEV-2016-0597, and MDM-2015-0509, some of which include ERDF funds from the European Union. IFAE is partially funded by the CERCA program of the Generalitat de Catalunya.  Research leading to these results has received funding from the European Research Council under the European Union's Seventh Framework Program (FP7/2007-2013) including ERC grant agreements 240672, 291329, and 306478.  We  acknowledge support from the Brazilian Instituto Nacional de Ci\^enciae Tecnologia (INCT) do e-Universo (CNPq grant 465376/2014-2).

This manuscript has been authored by Fermi Research Alliance, LLC under Contract No. DE-AC02-07CH11359 with the U.S. Department of Energy, Office of Science, Office of High Energy Physics.

\section*{Affiliations}
{\small
$^{1}$ Department of Astronomy, Shanghai Jiao Tong University, Shanghai 200240, China\\
$^{2}$ George P. and Cynthia Woods Mitchell Institute for Fundamental Physics and Astronomy, and Department of Physics and Astronomy, Texas A\&M University, College Station, TX 77843,  USA\\
$^{3}$ Observat\'orio Nacional, Rua Gal. Jos\'e Cristino 77, Rio de Janeiro, RJ - 20921-400, Brazil\\
$^{4}$ Fermi National Accelerator Laboratory, P. O. Box 500, Batavia, IL 60510, USA\\
$^{5}$ Department of Astronomy, University of Michigan, Ann Arbor, MI 48109, USA\\
$^{6}$ Department of Physics, University of Michigan, Ann Arbor, MI 48109, USA\\
$^{7}$ Astrophysics Research Centre, University of KwaZulu-Natal, Durban, 3696, South Africa\\
 $^{8}$ Kavli Institute for Cosmological Physics, University of Chicago, Chicago, IL 60637, USA\\
$^{9}$ Cerro Tololo Inter-American Observatory, NSF's National Optical-Infrared Astronomy Research Laboratory, Casilla 603, La Serena, Chile\\
$^{10}$ Laborat\'orio Interinstitucional de e-Astronomia - LIneA, Rua Gal. Jos\'e Cristino 77, Rio de Janeiro, RJ - 20921-400, Brazil\\
$^{11}$ Institute of Cosmology and Gravitation, University of Portsmouth, Portsmouth, PO1 3FX, UK\\
$^{12}$ CNRS, UMR 7095, Institut d'Astrophysique de Paris, F-75014, Paris, France\\
$^{13}$ Sorbonne Universit\'es, UPMC Univ Paris 06, UMR 7095, Institut d'Astrophysique de Paris, F-75014, Paris, France\\
$^{14}$ University Observatory, Faculty of Physics, Ludwig-Maximilians-Universit\"at, Scheinerstr. 1, 81679 Munich, Germany\\
$^{15}$ Department of Physics \& Astronomy, University College London, Gower Street, London, WC1E 6BT, UK\\
$^{16}$ Kavli Institute for Particle Astrophysics \& Cosmology, P. O. Box 2450, Stanford University, Stanford, CA 94305, USA\\
$^{17}$ SLAC National Accelerator Laboratory, Menlo Park, CA 94025, USA\\
$^{18}$ Instituto de Astrofisica de Canarias, E-38205 La Laguna, Tenerife, Spain\\
$^{19}$ Universidad de La Laguna, Dpto. Astrof\'isica, E-38206 La Laguna, Tenerife, Spain\\
$^{20}$ Center for Astrophysical Surveys, National Center for Supercomputing Applications, 1205 West Clark St., Urbana, IL 61801, USA\\
$^{21}$ Department of Astronomy, University of Illinois at Urbana-Champaign, 1002 W. Green Street, Urbana, IL 61801, USA\\
$^{22}$ Institut d'Estudis Espacials de Catalunya (IEEC), 08034 Barcelona, Spain\\
$^{23}$ Institute of Space Sciences (ICE, CSIC),  Campus UAB, Carrer de Can Magrans, s/n,  08193 Barcelona, Spain\\
$^{24}$ Jodrell Bank Center for Astrophysics, School of Physics and Astronomy, University of Manchester, Oxford Road, Manchester, M13 9PL, UK\\
$^{25}$ University of Nottingham, School of Physics and Astronomy, Nottingham NG7 2RD, UK\\
$^{26}$ Astronomy Unit, Department of Physics, University of Trieste, via Tiepolo 11, I-34131 Trieste, Italy\\
$^{27}$ INAF-Osservatorio Astronomico di Trieste, via G. B. Tiepolo 11, I-34143 Trieste, Italy\\
$^{28}$ Institute for Fundamental Physics of the Universe, Via Beirut 2, 34014 Trieste, Italy\\
$^{29}$ Hamburger Sternwarte, Universit\"{a}t Hamburg, Gojenbergsweg 112, 21029 Hamburg, Germany\\
$^{30}$ Centro de Investigaciones Energ\'eticas, Medioambientales y Tecnol\'ogicas (CIEMAT), Madrid, Spain\\
$^{31}$ Department of Physics, IIT Hyderabad, Kandi, Telangana 502285, India\\
$^{32}$ Jet Propulsion Laboratory, California Institute of Technology, 4800 Oak Grove Dr., Pasadena, CA 91109, USA\\
$^{33}$ Institute of Theoretical Astrophysics, University of Oslo. P.O. Box 1029 Blindern, NO-0315 Oslo, Norway\\
$^{34}$ Instituto de Fisica Teorica UAM/CSIC, Universidad Autonoma de Madrid, 28049 Madrid, Spain\\
$^{35}$ School of Mathematics and Physics, University of Queensland,  Brisbane, QLD 4072, Australia\\
$^{36}$ Santa Cruz Institute for Particle Physics, Santa Cruz, CA 95064, USA\\
$^{37}$ Center for Cosmology and Astro-Particle Physics, The Ohio State University, Columbus, OH 43210, USA\\
$^{38}$ Department of Physics, The Ohio State University, Columbus, OH 43210, USA\\
$^{39}$ Center for Astrophysics $\vert$ Harvard \& Smithsonian, 60 Garden Street, Cambridge, MA 02138, USA\\
$^{40}$ Australian Astronomical Optics, Macquarie University, North Ryde, NSW 2113, Australia\\
$^{41}$ Lowell Observatory, 1400 Mars Hill Rd, Flagstaff, AZ 86001, USA\\
$^{42}$ Department of Astrophysical Sciences, Princeton University, Peyton Hall, Princeton, NJ 08544, USA\\
$^{43}$ Instituci\'o Catalana de Recerca i Estudis Avan\c{c}ats, E-08010 Barcelona, Spain\\
$^{44}$ Institut de F\'{\i}sica d'Altes Energies (IFAE), The Barcelona Institute of Science and Technology, Campus UAB, 08193 Bellaterra (Barcelona) Spain\\
$^{45}$ Max Planck Institute for Extraterrestrial Physics, Giessenbachstrasse, 85748 Garching, Germany\\
$^{46}$ Department of Astronomy, University of California, Berkeley,  501 Campbell Hall, Berkeley, CA 94720, USA\\
$^{47}$ Institute of Astronomy, University of Cambridge, Madingley Road, Cambridge CB3 0HA, UK\\
$^{48}$ Department of Astronomy and Astrophysics, University of Chicago, Chicago, IL 60637, USA\\
$^{49}$ Department of Physics and Astronomy, University of Pennsylvania, Philadelphia, PA 19104, USA\\
$^{50}$ Department of Physics and Astronomy, Pevensey Building, University of Sussex, Brighton, BN1 9QH, UK\\
$^{51}$ Instituto de F\'isica, Pontificia Universidad Cat\'olica de Valpara\'iso, Casilla 4059, Valpara\'iso, Chile\\
$^{52}$ School of Physics and Astronomy, University of Southampton,  Southampton, SO17 1BJ, UK\\
$^{53}$ Computer Science and Mathematics Division, Oak Ridge National Laboratory, Oak Ridge, TN 37831\\
$^{54}$ Lawrence Berkeley National Laboratory, 1 Cyclotron Road, Berkeley, CA 94720, USA}\\

\appendix
\section{Background}
\label{sec:appendix}

For our measurement of the ICL profiles, we chose to estimate the background level based on the amount of light beyond 500\,kpc, as done in \citet{gol22}.  We chose this inner radial limit based on a comparison of background measurements starting between 500\,kpc and 1500\,kpc.  Figure~\ref{fig:Differences} highlights that as we change the inner limit of the background to larger radii, we continue to measure the same median Background level.  However, when the background starts at larger radii, there is a  significantly larger scatter in the background measurements, likely a result of the underlying noise present at large radii.  Additionally, while \citet{zha18} has measured the ICL out to radii of the order of 1Mpc, that was done using a stacked analysis, while this analysis looks at each cluster individually and we are thus unable to measure the ICL beyond 300\,kpc.

\renewcommand{\thefigure}{A\arabic{figure}}
\setcounter{figure}{0}
\begin{figure}
    \centering
    \includegraphics[width=10cm]{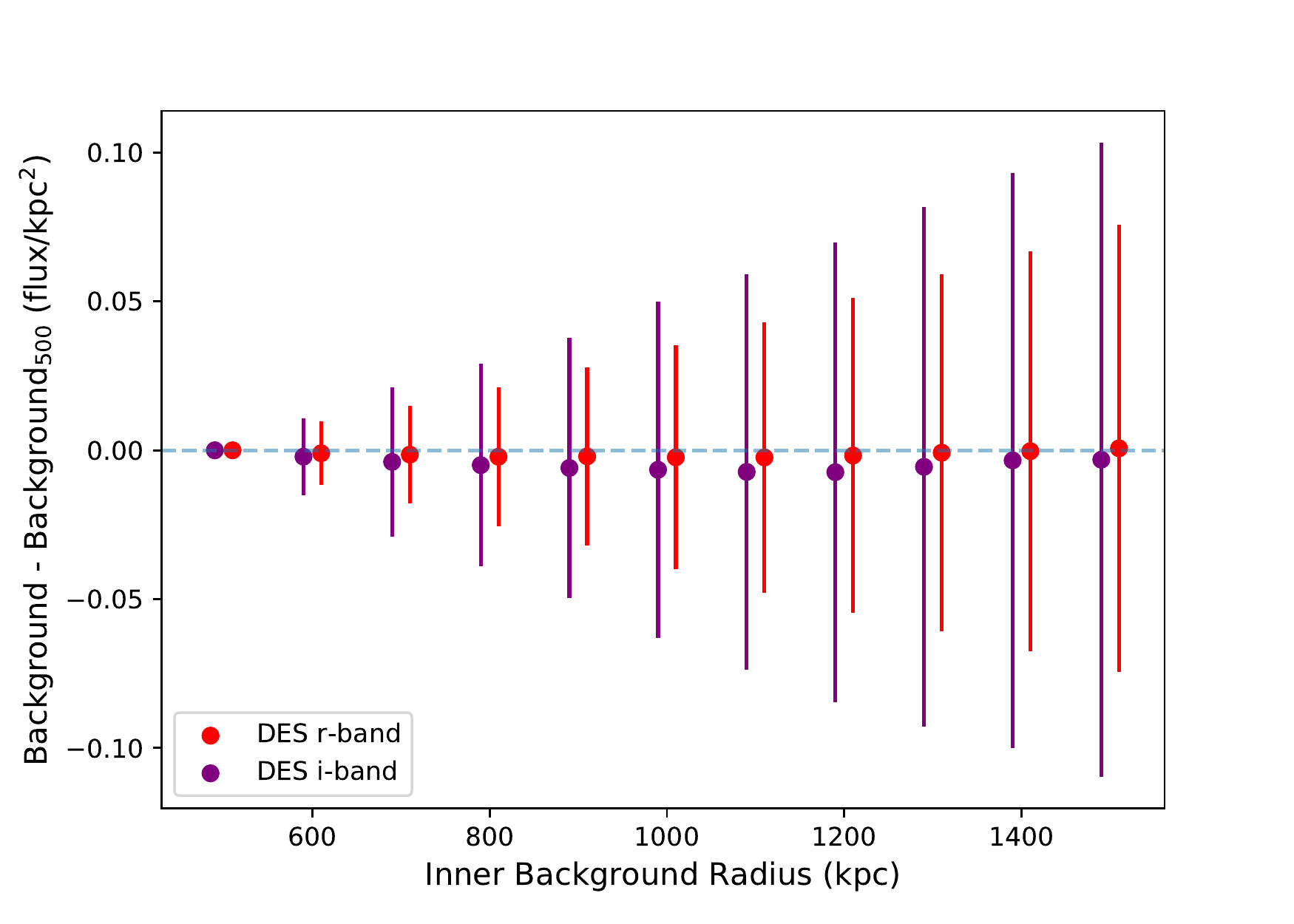}
    \caption{The difference in the background flux measurements (units of flux/kpc$^2$) for 11 unique radial ranges starting between 500\,kpc and 1500\,kpc and the 500\,kpc background spaced 100\,kpc apart.  The points represent the median value in the difference and the error bars the 1$\sigma$ standard deviation.  We see that the median value remains fixed while the scatter increases with radius.  Additionally, for our measurements, we use the DES selected zero point magnitude of 30.0.}    
    \label{fig:Differences}
\end{figure}

\section*{Data Availability}

The data underlying this article will be shared on reasonable request to the corresponding author.

\bibliographystyle{abc}

\label{lastpage}

\end{document}